\documentclass{article}
\usepackage[a4paper,top=2cm,bottom=2cm,left=2cm,right=2cm]{geometry}

\usepackage{authblk}
  \usepackage{paralist}
  \usepackage{graphics}
  \usepackage{epsfig} 
\usepackage{graphicx}

\usepackage{epstopdf}
 \usepackage[colorlinks=true]{hyperref}
\hypersetup{urlcolor=blue, citecolor=red}
\usepackage{amsmath} 
\usepackage{amsfonts}
\usepackage{bm}
\usepackage[bbgreekl]{mathbbol}
\DeclareSymbolFontAlphabet{\mathbbm}{bbold}
\DeclareSymbolFontAlphabet{\mathbb}{AMSb}
\usepackage{amssymb}
\usepackage{amsthm}
\usepackage{mathrsfs}

\usepackage{stackrel,graphicx}
\usepackage{enumitem}
\usepackage{cases}
\usepackage{booktabs}
\usepackage{braket}
\usepackage[strict]{changepage}
\usepackage{subfigure}
\usepackage[small,justification=justified]{caption}
\usepackage{titlesec} 
\usepackage{multicol}
\usepackage{hyperref}
\usepackage{chemarr}

\usepackage{cite}

\usepackage{amssymb,amsmath,amsthm}
\usepackage{booktabs}
\usepackage{multirow}
\usepackage[table]{xcolor}
\usepackage{subfigure}
\usepackage{chemfig}
\usepackage{mhchem}
\usepackage[right]{lineno}
\newcommand*{\myalign}[2]{\multicolumn{1}{#1}{#2}}

\newtheorem{oss*}{Observation}
\newtheorem*{rem*}{Remark}
\theoremstyle{definition}

\newcommand{\vb}{{\bf v}}

\newcommand{\x}{{\bf x}}

\newcommand{\ub}{{\bf b}}
\newcommand{\bV}{{\bf V}}
\newcommand{\bY}{{\bf Y}}

\def\T{\textrm{T}}
\providecommand{\keywords}[1]{\textbf{\textit{Keyworks --}} #1}

\DeclareMathOperator*{\argmaxA}{arg\,max}
\newenvironment{sistem}
{\left\lbrace\begin{array}{@{}l@{}}}
{\end{array}\right.}

\begin{document}

\title{Multi-scale modeling of Snail-mediated response to hypoxia in tumor progression}
\author[,1,2,3]{Giulia Chiari\footnote{Corresponding authors: martina.conte@polito.it, giulia.chiari@polito.it. These authors contributed equally to the work.}}
\author[*,1,4]{Martina Conte}
\author[1]{Marcello Delitala}
\affil[1]{\centerline{\small Department of Mathematical Sciences "G. L. Lagrange", Politecnico di Torino} \newline \centerline{\small Corso Duca degli Abruzzi 24, 10129 Torino, Italy}}
\affil[2]{\centerline {\small Department of Mathematics “G. Peano”, Università di Torino} \newline \centerline{\small  Via Carlo Alberto 10, 10124 Torino, Italy}}
\affil[3]{\centerline {\small Department of Mathematics, Swinburne University of Technology} \newline \centerline{\small  John St, Hawthorn VIC 3122, Australia}}
\affil[4]{\centerline {\small Division of Mathematical Oncology and Computational Systems Biology} \newline \centerline{\small Department of Computational and Quantitative Medicine, Beckman Research Institute} \newline \centerline{\small City of Hope National Medical Center, Duarte, CA, USA}}
\date{\today}                     
\setcounter{Maxaffil}{0}
\renewcommand\Affilfont{\itshape\small}
\maketitle

\begin{abstract}
Tumor cell migration within the microenvironment is a crucial aspect for cancer progression and, in this context, hypoxia has a significant role. An inadequate oxygen supply acts as an environmental stressor inducing migratory bias and phenotypic changes. In this paper, we propose a novel multi-scale mathematical model to analyze the pivotal role of Snail protein expression in the cellular responses to hypoxia. Starting from the description of single-cell dynamics driven by the Snail protein, we construct the corresponding kinetic transport equation that describes the evolution of the cell distribution. Subsequently, we employ proper scaling arguments to formally derive the equations for the statistical moments of the cell distribution, which govern the macroscopic tumor dynamics. Numerical simulations of the model are performed in various scenarios with biological relevance to provide insights into the role of the multiple tactic terms, the impact of Snail expression on cell proliferation, and the emergence of hypoxia-induced migration patterns. Moreover, quantitative comparisons with experimental data show the model's reliability in measuring the impact of Snail transcription on cell migratory potential. Through our findings, we shed light on the potential of our mathematical framework in advancing the understanding of the biological mechanisms driving tumor progression.
\end{abstract}
 

\section{Introduction}\label{sec:intro}
Migration of tumor cells into the normal tissue under the influence of biochemical and biophysical components of the micro-environment is one of the hallmarks of cancer \cite{hanahan2011hallmarks}. However, because of the highly complex biology at the cellular and molecular level and in the interactions with the surrounding environment, the exact dynamics driving cell migration are still not completely well understood. Tissue oxygenation is one of the prominent traits in this context. It has been suggested that oxygen concentration highly influences the switch between migrating and proliferating cell behavior, and the invasiveness and aggressiveness of the tumor cells. Moreover, the deprivation of oxygen acts as an environmental stressor, promoting a long series of mutations that strongly impact the tumor dynamics \cite{martinez2012hypoxic}.\\ 
\indent Hypoxia, defined as an insufficient supply of oxygen, has long been recognized as a contributing factor to the tumor microenvironment (TME) \cite{chen2023hypoxic}. Not only it can induce a pronounced migratory bias of the cells towards favorable areas, but it can even determine their phenotype and interaction strategies. It has been clinically observed that, in solid tumors, the oxygen distribution is heterogeneous, with oxygen levels ranging from normal to hypoxic and severe hypoxic \cite{martinez2012hypoxic,conte2024emergence}. Under hypoxic conditions, tumor cells undergo morphological and molecular changes to adjust their behaviors and acquire the abilities to adapt to hypoxia and escape apoptosis \cite{ruan2009role}. There are several adaptive responses of tumor cells to hypoxia \cite{chen2023hypoxic,conte2024emergence}, which may involve the secretion of specific transcription factors, like hypoxia-inducible factor 1 (HIF-1), the upregulation of hypoxia-inducible angiogenic factors, sustaining new vessels formation \cite{yehya2018angiogenesis}, or the glycolysis activation \cite{samanta2018metabolic}. Among others, HIF-1 activation controls the expression of Snail transcription. \\
\indent The Snail superfamily of transcription factors includes Snail-1, Slug, and Scratch proteins \cite{kaufhold2014central}. It is well documented that Snail protein directly represses E-cadherin and, thus, it is a key inducer of epithelial-mesenchymal transition, a biological process defining progression from a polarised epithelial cell phenotype to a mesenchymal phenotype \cite{peinado2004snail,hotz2007epithelial}. In addition to regulating epithelial-to-mesenchymal transition (EMT) and cell migration, overexpression of Snail induces resistance to apoptosis and tumor recurrence \cite{jin2010snail}. The Snail-mediated survival of epithelial cells may thus enhance the ability of tumor cells to invade and metastasize. Overexpression of Snail has been reported to be a sufficient inducer of EMT as well as a predictor for an aggressive tumor phenotype. It has recently been demonstrated that Snail expression is induced by hypoxic conditions and is regulated by HIF-1$\alpha$ expression at the transcriptional level \cite{cano2000transcription, lundgren2009hypoxia,imai2003hypoxia}. Up-regulation of Snail-1 correlates with metastasis and poor prognosis, whereas silencing of Snail-1 is critical for reducing tumor growth and invasiveness \cite{barrallo2005snail}. Since such complex processes and their mutual conditioning scenarios are difficult to assess experimentally, mathematical models can help understand the underlying biological mechanisms, test hypotheses, and even make predictions.\\
\indent Cancer progression is a complex process, involving several factors, and taking place as both an individual and a collective process. Microscopic intracellular dynamics, occurring at the individual cell level, influence the mesoscopic behavior of the cell population, which determines the macroscopic evolution for their density. Previous models for tumor invasion have been proposed in discrete or continuous frameworks. The former are based on the description of individual cell dynamics moving on a lattice \cite{giverso2010individual,szabo2013cellular} and can involve continuous equations for the evolution of external factors (e.g., chemoattractant concentration, density of ECM fibers, low pH levels) - the so-called hybrid models \cite{jeon2010off,anderson2007hybrid,anderson2009microenvironment}. Concerning the latter, different classes of fully continuous models for tumor cell migration have been developed. Many of them are versions or extensions of the classical reaction-diffusion model proposed by Murray \cite{murray1989introduction}. More recent works take into account the advection bias of tumor cells describing motility adjustment to extracellular signals. Some of these are directly set on the macroscopic scale and rely on balance equations for mass, flux, or momentum \cite{ambrosi2002closure,chaplain2019derivation,anderson2000mathematical}, or on integro-differential equations accounting for the development of specific intra-tumor structure \cite{ardavseva2020mathematical,bitsouni2018non,chiari2023hypoxia, fiandaca2021mathematical,strobl2020mix,szymanska2024mathematical} (see also the review in \cite{kolbe2021modeling} for settings with multiple taxis in the larger context of cell migration). More recently, the use of kinetic transport equations (KTEs) in the kinetic theory of active particles (KTAP) framework has been largely applied to the study of cell migration, in general, \cite{chalub2004kinetic,conte2022multi,hillen2006m5,loy2020kinetic,chauviere2007modeling,conte2023non,kelkel2012multiscale}, and in the specific context of tumor evolution \cite{conte2020glioma,engwer2015glioma,buckwar2023stochastic,engwer2016effective,kolbe2021modeling,lorenz2014class,zhigun2022novel}. Kinetic models are intrinsically multi-scale models that characterize the dynamics of distribution functions of tumor cells which may depend, besides time and position, on several kinetic variables, such as microscopic velocity or activity variables. These models use Boltzmann-type equations for the cell population density and scaling arguments to derive the macroscopic setting. Among those models, \cite{conte2023mathematical,conte2021mathematical,engwer2017structured,corbin2021modeling,kumar2021multiscale,dietrich2022multiscale} accounts for effects of hypoxia or hypoxia-driven acidity on the migration and invasion process of tumor cells. Concerning the tactic terms obtained in this class of models, in \cite{conte2023mathematical,engwer2017structured,engwer2015glioma,dietrich2022multiscale} they are derived from the description of subcellular dynamics for receptor binding, which leads to transport terms w.r.t. the activity variables in the mesoscopic KTE. In \cite{conte2021mathematical,kumar2021multiscale,loy2020kinetic}, instead, the use of turning rates depending on the pathwise gradient of some chemotactic signal leads to various types of taxis at the macroscopic level. Moreover, in \cite{chauviere2007modeling,corbin2021modeling,dietrich2022multiscale}, forces and stress, acting on the cells and depending on the chemical and physical composition of the environment, translates into transport terms w.r.t. the velocity variable in the corresponding KTE. In particular, in \cite{dietrich2022multiscale}, the authors consider a flux-limited description of the transport terms. Flux limitations have been introduced in the modeling of cell motility to reduce the infinite speed of propagation triggered by linear diffusion and the excessive influence of the latter on the spread of cells. Their derivation from KTEs has been provided formally in \cite{bellomo2010multiscale} and rigorously in \cite{perthame2019flux}. In both cases, the derivation is based on an appropriate choice of the signal response function involved in the turning operator and depends on the directional derivative of the signal. In \cite{dietrich2022multiscale}, instead, an alternative approach based on characterizing velocity dynamics at the single-cell level is proposed. Lastly, concerning the mathematical modeling of Snail dynamics, several works have been proposed for theoretically studying Snail's role in the epithelial-to-mesenchymal transition process, especially looking at the interactions among microRNAs and transcription factors - miR-34, miR-200, Zeb, and Snail - at the single-cell level \cite{jia2017distinguishing,lu2013tristability,tripathi2021mathematical}, while its connection to cellular motility has not been largely investigated. \\ 
\indent In the present work, we propose a multi-scale mathematical modeling approach for describing tumor invasion in response to tissue hypoxia, investigating the interplay between molecular signaling pathways and cell dynamics. The model connects single-cell behaviors driven by Snail expression with macroscopic scale dynamics describing tumor migration in the tissue. Starting from the approach proposed in \cite{dietrich2022multiscale}, we introduce a novel description of the internal variable dynamics for Snail expression and we account for flux-limited operators in the single-cell velocity dynamics. At the mesoscopic level, cell evolution is described in terms of a classical kinetic transport equation with different formulations for the proliferative operator. From this description, using proper up-scaling arguments, we derive the macroscopic setting. Moreover, we numerically investigate the model's capability of capturing different biologically relevant scenarios concerning hypoxia and Snail effects on cell migration and proliferation. We show how the model is reliable in replicating different experimental results and offers new perspectives for interpreting experimental findings.\\
\indent The paper is organized as follows. Section \ref{model} provides the setup of microscopic and mesoscopic equations for the dynamics of tumor cells. Section \ref{Macro_der} contains the derivation of the macroscopic equation for tumor cell density evolution, which features flux-limited chemotaxis towards increasing oxygen concentrations and self-diffusion, as well as a proliferation term modeling the inverse correlation between moving and proliferating cell capability in two possible manner. In Section \ref{Sec_numsim}, four numerical experiments are proposed to show both the qualitative behavior of the model in different scenarios and its capability to quantitatively replicate experimental data concerning Snail impact of tumor cell migration in different oxygen conditions. Finally, Section \ref{sec_conclusion} provides a discussion of the main outcomes of our model, along with some perspectives.

\section{Modeling}\label{model}
 In this note, we propose a multi-scale model for describing tumor progression in response to tissue hypoxia, whose influence on the cancer cells is mediated by Snail dynamics. Following the well-established literature regarding multi-scale models for tumor invasion \cite{kelkel2012multiscale,engwer2015glioma,engwer2017structured,corbin2021modeling,conte2020glioma}, the model setting proposed here is built using the classical tools and methods of kinetic theory. Our main aim is to obtain a detailed description of the tumor cell dynamics, taking into account the effect of microscopic signaling pathways in the mechanisms of tumor response to hypoxia. \\ 
 \indent Starting from the microscopic level of interaction between cells and oxygen, we consider the dynamics of the Snail signaling pathway, which is involved in the cell response to hypoxic microenvironmental conditions. Moreover, following \cite{dietrich2022multiscale}, we provide a microscopic description of velocity dynamics, which depend on oxygen and macroscopic cell density tactic gradient, with Snail expression influencing cell motility. Then, we set up the corresponding kinetic transport equation describing the evolution of the cell distribution
 in relation to the prescribed microscopic dynamics.

\subsection{Microscopic scale}\label{cell level}
At this level, we model the dynamics of the microscopic variable $y\in\bY=(0,y_0)$, describing the expression of the Snail protein, and the microscopic cell velocity $\vb\in\bV$, both influenced by the oxygen levels. We assume oxygen to be time-independent, i.e., $O_2=O_2(\x)$ has a fixed distribution that does not evolve in time. Possible extensions of this approach are discussed in Section \ref{sec_conclusion}. 

\subsubsection{Dynamics of Snail protein expression}
\noindent Concerning Snail protein expression, we model the process of protein synthesization from gene transcription and its regulation depending on the oxygen dynamics. Snail proteins are involved in the regulation of hypoxia-driven cell migration and invasion \cite{imai2003hypoxia,zhu2013hypoxia}. It has been shown in several different types of human cancer that overexpression of Snail induces invasion and metastasis \cite{blanco2002correlation,roy2005transcriptional,poser2001loss,zhu2013hypoxia}. The expression of the Snail protein is controlled by the oxygen levels, decreasing when the tumor mass is properly oxygenated. Relying on the description of the temporal evolution of the total level of Snail proposed in \cite{tripathi2021mathematical}, here we model its dynamics with the following equation:
\begin{equation}
\dfrac{dy}{dt}=g_sH(y)H(O_2)-\gamma_s y\,,
\label{micro_y}
\end{equation}
where $g_s$ and $\gamma_s$ represent the basal transcription and degradation rates, respectively, while the functions $H(\cdot)$ model the transcription activation/inhibition mechanisms. We recall here that $O_2=O_2(\x)$. Generally, the functions $H(A)$ can be described as
\[
H(A)=\dfrac{1+\lambda_{A}\frac{A}{A_{0}}}{1+\frac{A}{A_{0}}},
\]
where $A$ is a generic agent influencing the transcription and $A_0$ is its reference value. In particular, $\lambda_{A}>1$ models an activation mechanism supported by the agent $A$, while $\lambda_{A}<1$ refers to an inhibition mechanism driven by $A$. In the case of Snail, it has been shown that it has a self-regulatory (inhibition) mechanism \cite{de2010snail,lu2013microrna}, which we describe as
\[
H(y)=\dfrac{1}{1+\frac{y}{y_0}}\,,
\] 
with $y_0$ maximum Snail expression.
\begin{oss*}
We assume $\lambda_A=0$ to avoid overloading the model with further parameters. However, other choices of this function may be considered. 
\end{oss*}
\noindent Concerning $O_2$, it exerts an inhibitory mechanism on the transcription of Snail. However, for the definition of the microscopic dynamics and the derivation of the macroscopic model, we keep a general expression for the function $H(O_2)$. We then specify the value of the parameter $\lambda_{O_2}$ used for the numerical simulations in Table \ref{parameter_const}. Rescaling $y/y_0 \rightsquigarrow y$ and $g_s/y_0 \rightsquigarrow g_s$, we simplify the notation as
\begin{equation}
\dfrac{dy}{dt}=g_s\dfrac{1}{1+y}H(O_2)-\gamma_s y:=G(y,O_2)
\label{micro_y2}
\end{equation}
with $y\in\bY=(0,1)$. Looking at its steady state $y^*$, we observe that
\begin{equation*}
\begin{split}
\dfrac{dy}{dt}=0&\Longleftrightarrow g_s\dfrac{1}{1+y}H(O_2)-\gamma_s y=0\\[0.2cm]
&\Longleftrightarrow \left[g_sH(O_2)-\gamma_s y(1+y)\right]\dfrac{1}{1+y}=0\\[0.2cm]
&\Longleftrightarrow y(1+y)=\dfrac{g_s}{\gamma_s}H(O_2)\\[0.2cm]
&\Longleftrightarrow y^2 +y-\dfrac{g_s}{\gamma_s}H(O_2)=0\\[0.2cm]
&\Longleftrightarrow y=-\dfrac{1}{2}\left(1\pm\sqrt{1+4\dfrac{g_s}{\gamma_s}H(O_2)}\right)\,.
\end{split}
\end{equation*}
As $y$ represents a biological quantity accounting for the expression of Snail, no negative values are admitted. Thus, the only acceptable steady-state solution is given by
\[
y^*=\dfrac{1}{2}\left(\sqrt{1+4\dfrac{g_s}{\gamma_s}H(O_2)}-1\right)\,.
\]
\begin{oss*}\label{oss_ystar}
To ensure that the equilibrium distribution $y^*$ belongs to $\bY$, we have to require that
\begin{equation}\label{cond_y}
    \dfrac{g_s}{\gamma_s} \max_{\x \in\mathbb{R}^n}\{H(O_2(\x))\}< 2\,.
\end{equation} If we consider an inhibitory function $H(O_2)$ such that $\lambda_{O_2}=0$, than the condition reads $g_s< 2\gamma_s$.
\end{oss*}

\subsubsection{Dynamics of cell velocity}
Concerning the microscopic velocity $\vb$, we model the mechanism by which cells tend to migrate by aligning to two different gradients. Precisely, increasing gradients of oxygen attract tumor cells towards better oxygenated areas, while tumor cells tend to avoid crowded regions with high cell densities. In both cases, the smaller the amount of Snail expression, the lower the cell's tendency to move along the directions of these gradients. Under these assumptions, velocity dynamics are modeled with the following equation
\begin{equation}
\dfrac{d\vb}{dt}=g(y,O_2,M)-a_2\vb\,,
\label{micro_v}
\end{equation}
where the function $g(y,O_2,M)$ describes cell acceleration, while the second term on the right-hand-side models cell deceleration, with $a_2$ a positive scaling constant. Cells, in fact, tend to slow down or randomly move in the absence of external signals. Concerning cell acceleration, we set
\begin{equation}\label{micro_g}
g(y,O_2,M)=a_1\ub(y,O_2,M)\,.
\end{equation}
Here, $\ub(y,O_2,M)$ is the vector gradient modeling cell alignment along the directions given by the gradient of oxygen $O_2$ and the gradient of macroscopic tumor cell density $M$, while $a_1$ is a positive scaling constant. As introduced above, we assume that the cell's tendency to follow oxygen gradient is enhanced by high Snail expression, which is one of the mechanisms of cell response to hypoxia. At the same time, since low levels of Snail promote high levels of E-cadherin expression \cite{wang2015snail,kaufhold2014central}, which is responsible for cell-cell adhesion, we assume that cell tendency to avoid high cell density region is also positively regulated by $y$. In fact, high levels of Snail would promote less adhesion between cells and, thus, a more enhanced tendency to escape from the tumor core. Thus, we choose
\begin{equation}\label{b_fun}
\ub(y,O_2,M)=y\,\left(\beta\dfrac{\nabla_\x O_2}{\sqrt{\left(\dfrac{O_{2,0}}{X}\right)^2+|\nabla_\x O_2|^2}}-(1-\beta)\dfrac{\nabla_\x M}{\sqrt{\left(\dfrac{K_M}{X}\right)^2+|\nabla_\x M|^2}}\right)\,.
\end{equation} 

Here, $K_M$ and $O_{2,0}$ are reference values for tumor cells and oxygen, $X>0$ is a constant to be selected in correspondence to appropriate time and length scales, while the parameter $\beta\in[0,1]$ weights the contributions of the two tactic terms. Equation \eqref{micro_v} can be written as
\begin{equation}
\dfrac{d\vb}{dt}=a_1y\,\left(\beta\dfrac{\nabla_\x O_2}{\sqrt{\left(\dfrac{O_{2,0}}{X}\right)^2+|\nabla_\x O_2|^2}}-(1-\beta)\dfrac{\nabla_\x M}{\sqrt{\left(\dfrac{K_M}{X}\right)^2+|\nabla_\x M|^2}} \right)-a_2\vb=:S(\vb,y,O_2,M)\,.
\label{micro_v2}
\end{equation}
\noindent We observe that $g(y,O_2,M)$ is bounded
\[
|g(y,O_2,M)|=\left|a_1\,\ub\right|<a_1
\]
and, thus, the speed $s=|\vb|<\dfrac{a_1}{a_2}:=s_{ub}$, with $s_{ub}$ an upper bound for cell speed. Finally, we complete the microscopic level system by modeling the changes in the cell position $\x\in\mathbb{R}^n$ as
\begin{equation}
\dfrac{d\x}{dt}=\vb\,.
\label{micro_x}
\end{equation}
Collecting equations \eqref{micro_y2}, \eqref{micro_v2}, and \eqref{micro_x}, the complete system for the microscopic level dynamics reads
\begin{equation}
\begin{sistem}
\dfrac{d\x}{dt}=\vb\,,\\[.2cm]
\dfrac{d\vb}{dt}=a_1\,y\,\left(\beta\dfrac{\nabla_\x O_2}{\sqrt{\left(\dfrac{O_{2,0}}{X}\right)^2+|\nabla_\x O_2|^2}}-(1-\beta)\dfrac{\nabla_\x M}{\sqrt{\left(\dfrac{K_M}{X}\right)^2+|\nabla_\x M|^2}}\right)-a_2\vb\,,\\[1cm]
\dfrac{dy}{dt}=g_s\dfrac{1}{1+y}H(O_2)-\gamma_s y\,.
\end{sistem}
\end{equation}

\subsection{Mesoscopic scale}
At this level, we consider the cell density distribution $p(t,\x,\vb,y):[0,T]\times\mathbb{R}^n\times\bV\times\bY\to\mathbb{R}$, depending on time $t>0$, position $\x\in\mathbb{R}^n$, microscopic velocity $\vb\in\bV$, and internal variable for Snail protein expression $y\in\bY$. In particular, the microscopic velocity vector $\vb$ can be written as $\vb=s{\bm \vartheta}$ with cell speed $s\in(0,s_{ub})$ and cell direction ${\bm \vartheta}\in\mathbb{S}^{n-1}$. For describing the mesoscopic dynamics of tumor cells, we consider the following kinetic transport equation:
\begin{equation}
\dfrac{\partial p}{\partial t} +\nabla_\x\cdot\,(\vb p) +\dfrac{\partial }{\partial y}(G(y,O_2)p)+\nabla_\vb\cdot\,(S(\vb,y,O_2,M) p)=\mathcal{P}[p]\,.
\label{peq}
\end{equation}
 Here, the functions $G(y,O_2)$ and $S(\vb,y,O_2,M)$ are given by \eqref{micro_y2} and \eqref{micro_v2}, respectively, while the operator $\mathcal{P}[p]$ accounts for the proliferation process. We generally describe it as
 \[
 \mathcal{P}[p]=\mu_1(M,O_2,s)\int_\bY\mu_2(y')\chi(t,\x,y,y')p(t,\x,\vb,y')dy'\,.
 \]
 Here, the coefficient function $\mu_1(M,O_2,s)$ accounts for the possible effect of cell speed and oxygen level changes, as well as overcrowding of the environment, on cell proliferation. Instead, the integral operator, involving the coefficient function $\mu_2(y)$ and the kernel $\chi(t,\x,y,y')$, describes the role of Snail expression in the proliferation process. Precisely, the kernel $\chi(t,\x,y,y')$ represents the probability for a daughter cell of receiving a Snail expression $y$ after the division of a mother cell with Snail expression $y'$. We propose two possible expressions for the proliferation term, both based on the assumption that cell capabilities of moving and proliferating are inversely correlated (go-or-grow hypothesis \cite{zheng2009cell} and tradeoffs \cite{gallaher2019impact}). 
 \begin{itemize}
     \item[$\bullet$] In the first case, we set 
\[
\mu_1(M,O_2,s):=\mu\dfrac{s_{ub}-s^*}{s_{ub}}\left(1-\dfrac{M}{K_M}\right)\dfrac{O_2}{O_{2,0}+O_2}\qquad\text{and}\qquad\mu_2(y):=1 \,,\,\,\,\forall y\in\bY.
\]
Moreover, we assume that the level of Snail expression in a daughter cell is equal to the one of its mother, i.e., $\chi(t,\x,y,y')=\delta(y-y')$. With these assumptions, the proliferative operator reads 
\begin{equation}\label{P1}
\mathcal{P}_1[p]=\mu\dfrac{s_{ub}-s^*}{s_{ub}}\left(1-\dfrac{M}{K_M}\right)\dfrac{O_2}{O_{2,0}+O_2}p(t,\x,\vb,y)
\end{equation}
Here, the impact of proliferation-migration tradeoffs is captured by the inverse relationship between $\mu_1(M,O_2,s)$ and cell speed. The influence of $y$ is indirectly incorporated, namely through its impact on the variation of $s$. As the adaptation of cell speed to the surrounding environment occurs more rapidly than proliferation, the cell speed is approximated by the module of the steady-state velocity, denoted as $s^*=|\vb^*|$.
\item[$\bullet$] In the second case, we set  
\[
\mu_1(M,O_2,s):=\mu\left(1-\dfrac{M}{K_M}\right)\dfrac{O_2}{O_{2,0}+O_2}\qquad\text{and}\qquad\mu_2(y):=1-y \,,\,\,\,\forall y\in\bY.
\]
With this choice, the mentioned tradeoffs are taken into account in the integral term, which models a reduced proliferation for high levels of Snail expression. Moreover, we assume that the kernel does not depend on the level of Snail expression of the mother cell, i.e., $\chi=\chi(t,\x,y)$, and distribution of the level of Snail expression in a daughter cell is symmetrical around the steady-state $y^*$, i.e.,
\[
\int_\bY(y-y^*)\chi(t,\x,y)dy=0\,.
\]
With these assumptions, the proliferative operator reads 
\begin{equation}\label{P2}
\mathcal{P}_2[p]:=\mu\left(1-\dfrac{M}{K_M}\right)\dfrac{O_2}{O_{2,0}+O_2}\int\limits_\bY(1-y')\chi(t,\x,y)p(t,\x,\vb,y')dy'
\end{equation}
 \end{itemize}
In both cases, the macroscopic cell density $M$ is defined by 
\begin{equation}\label{M_def}
    M:=\int\limits_\bV\int\limits_\bY p(t,\x,\vb,y)dy d\vb\,.
\end{equation}
The introduced descriptions of the proliferation mechanism are such that proliferation is reduced for highly motile cells, with a direct or indirect effect of Snail expression on the tradeoff. In both cases, the parameter $\mu$ represents the constant tumor proliferation rate. We include a direct effect of oxygen on tumor cell proliferation, which is enhanced at higher oxygen levels. We compare the effects of these two descriptions of cell proliferation on the macroscopic cell dynamics in the numerical {\bf Experiment 2} in Section \ref{Sec_numTest2}. 

\section{Derivation of macroscopic system}\label{Macro_der}
Due to the high dimension of \eqref{peq}, solving directly this kinetic equation has to face several challenges, especially related to its complexity and a high computational cost. Therefore, in this Section, we aim at deducing a macroscopic counterpart of \eqref{peq}.  Performing a proper model upscaling, we obtain the equations for the statistical moments of the cell distribution. These describe the dynamics of tumor cells, which are driven by limited diffusion and oxygen-mediated drift, and the evolution of the average Snail expression of the tumor population.

\subsection{Assumptions}\label{Sec_assum} 
To obtain a closed system of macroscopic equations from the integration of \eqref{peq} w.r.t $y$ and $\vb$, we need to make the following assumptions on the moments of the distribution function:
\begin{equation*}
\nabla_\x \cdot \int\limits_{\bV}(\vb-\vb^*)(v_i-v_i^*)pd\vb\approx 0\,,\qquad\quad  \int\limits_{\bV}\int\limits_{\bY}(v_i-v_i^*)(y-y^*)pdyd\vb\approx 0\,,\qquad\quad \int\limits_{\bV}\int\limits_{\bY}(y-y^*)^2pdyd\vb\approx 0\,.
\end{equation*} 
Precisely, we indicate with $v_i$ the $i$-th component of the velocity vector $\vb$, while with $y^*$ and $\vb^*$ the steady-states of the microscopic equations \eqref{micro_y2} and \eqref{micro_v2}, respectively. We assume that some of the second-order moments of the tumor cell distribution w.r.t. deviations of $\vb$ and $y$ from their steady-states are negligible, as well as the second-order moment w.r.t $y$. These are reasonable choices since the microscopic dynamics of protein expression and velocity changes happen faster in comparison to the behavior of tumor cells that share the same regimes of activity and velocity variables.\\
\indent We recall that $\bY=(0,1)$ and $\bV=B_{s_{ub}}^n(0)=(0,s_{ub})\times\mathbb{S}^{n-1}$. Following the approach proposed in \cite{dietrich2022multiscale,conte2020glioma,conte2021mathematical,engwer2015glioma}, we assume 
the distribution $p$ to be compactly supported in the $\bV\times\bY$ space. Precisely, for equation \eqref{peq}, boundary conditions w.r.t. these variables need to be prescribed at the inflow boundary of $\bY$ and $\bV$. Considering the dynamics in \eqref{micro_y2}, a protein expression state $y \in \partial \bY$ is part of the inflow boundary if $G(y,O_2)\cdot {\bf n}\le 0$ for ${\bf n}$ outward normal on the boundary. Given $\partial \bY=\{0,1\}$, it holds
\begin{equation}\label{BC_Y}
\begin{split}
 &G(0,O_2)\cdot n=g_sH(O_2)\cdot (-1)<0\\[0.3cm]
&G(1,O_2)\cdot n=\left(2\,g_sH(O_2)-\gamma_s\right)\cdot (1)\le 0\,\,\, \text{if condition \eqref{cond_y} holds.} 
 \end{split}
\end{equation}
Thus, the inflow boundary of $\bY$ coincides with $\partial \bY$ and boundary conditions can be prescribed on the whole $\partial \bY$. Instead, considering the dynamics in \eqref{micro_v2} and a velocity vector $\vb \in \partial \bV$, we have that $|\vb|=s_{ub}$ and the outward normal ${\bf n}=\vb/s_{ub}$. The velocity vector $\vb$ is part of the inflow boundary if $S(\vb,y,O_2,M)\cdot {\bf n}\le 0$, i.e.,
\begin{equation}\label{BC_V}
\begin{split}
  S(\vb,y,O_2,M)\cdot {\bf n}&=\dfrac{a_1}{s_{ub}} \left( {\bf b}(y,O_2,M)\cdot \vb\right) -\dfrac{a_2}{s_{ub}}\left( \vb\cdot \vb\right)\\[0.3cm]
  &\le\dfrac{a_1}{s_{ub}}|{\bf b}(y,O_2,M)|\,|\vb|-\dfrac{a_2}{s_{ub}}|\vb|^2\\[0.3cm]
  &\le a_1-a_2 s_{ub}=0\,.
  \end{split}
\end{equation}
Thus, the inflow boundary of $\bV$ coincides with $\partial \bV$ and boundary conditions can be prescribed on the whole $\partial \bV$. Precisely, \eqref{BC_Y} and \eqref{BC_V} allow to conclude that the characteristics of the transport part of equation \eqref{peq} that start in $\mathbb{R}^n\times \bV \times \bY$ do not leave this set.
 
\subsection{Equation for the moments}
To derive the macroscopic equations we first rescale the quantities introduced above as $p/K_M \rightsquigarrow p$, $M/K_M \rightsquigarrow M$, $O_2/O_{2,0} \rightsquigarrow O_2$, and $s^*/s_{ub} \rightsquigarrow s^*$. Then, we introduce a small parameter $\varepsilon\ll 1$ to rescale time and space as
\begin{equation*}
\begin{split}
&\hat{t}=\varepsilon^\kappa t\,,\quad \hat{\x}=\varepsilon \x\,,\\[0.3cm]
&\hat{g}_s=\varepsilon^{-\nu}g_s\,,\quad\hat{\gamma}_s=\varepsilon^{-\nu}\gamma_s\,,
\end{split}
\end{equation*}
with $\kappa, \nu >0$. The rescaling of the reaction rates $g_s$ and $\gamma_s$ with a negative power of $\varepsilon$ is chosen to reflect the fact that these dynamics are the fastest among all included processes. Moreover, we assume that $1/X$ is of order $\varepsilon$, and the proliferation rate scales with $\varepsilon^\kappa$. For simplicity of writing, we drop the hat symbol from all variables and, thus, equation \eqref{peq} reads
\begin{equation}\label{p_mes_rescale}
\varepsilon^\kappa\dfrac{\partial p}{\partial t} +\varepsilon\nabla_\x\cdot\,(\vb p) +\varepsilon^{-\nu}\dfrac{\partial }{\partial y}(G(y,O_2)p)+\nabla_\vb\cdot\,(S(\vb,y,O_2,M)p)=\varepsilon^\kappa \mathcal{P}_k[p]
\end{equation}
where 
\begin{equation*}
\begin{split}
&G(y,O_2)=g_s\dfrac{1}{1+y}{H}(O_{2})-\gamma_sy\,,\\[0.2cm]
&S(\vb,y,O_2,M)=a_1y\,\left(\beta\dfrac{\nabla_\x O_2}{\sqrt{1+|\nabla_\x O_2|^2}}-(1-\beta)\dfrac{\nabla_\x M}{\sqrt{1+|\nabla_\x M|^2}}\right)-a_2\vb\,,\\[0.4cm]
\end{split}
\end{equation*}
and $\mathcal{P}_k[p]$ would be given by either
\begin{equation}\label{P1_res}
\mathcal{P}_1[p]=\mu(1-s^*)\left(1-M\right)\dfrac{O_2}{1+O_2}p(t,\x,\vb,y)
\end{equation}
or
\begin{equation}\label{P2_res}
\mathcal{P}_2[p]:=\mu\left(1-M\right)\dfrac{O_2}{1+O_2}\int\limits_\bY(1-y')\chi(t,\x,y)p(t,\x,\vb,y')dy'\,,
\end{equation}
which corresponds to the rescaled versions of \eqref{P1} or \eqref{P2}, respectively. Together with the introduced macroscopic tumor density \eqref{M_def}, let consider the following notations for the moment of the distribution function $p$:

\begin{align*}
&m(t,\x,y):=\int\limits_\bV p(t,\x,\vb,y)d\vb\,,\qquad\qquad \quad \,\,\,m^\vb(t,\x,y):=\int\limits_\bV \vb p(t,\x,\vb,y)d\vb\,, \\[0.3cm]
&m_i^\vb(t,\x,y):=\int\limits_\bV v_i p(t,\x,\vb,y)d\vb\,,
\qquad\quad\,\,\,\,\,\,\,  M^\vb_i(t,\x)=\int\limits_\bY\int\limits_\bV v_i p(t,\x,\vb,y)d\vb dy\,,\\[0.3cm]
&M^\vb(t,\x)=\int\limits_\bY\int\limits_\bV\vb p(t,\x,\vb,y)d\vb dy\,,\qquad\quad\,\,M^y(t,\x)=\int\limits_\bY\int\limits_\bV yp(t,\x,\vb,y)d\vb dy\,. 
\end{align*}

\noindent Considering \eqref{p_mes_rescale} and integrating w.r.t $\vb$ we obtain
\begin{equation}
\varepsilon^\kappa\dfrac{\partial m}{\partial t} +\varepsilon\nabla_\x \cdot\,m^\vb +\varepsilon^{-\nu}\dfrac{\partial }{\partial y}(G(y,O_2)m)+\int\limits_\bV \nabla_\vb\cdot\,S(\vb,y,O_2,M) p d\vb=\varepsilon^\kappa\int\limits_\bV\mathcal{P}_k[p]d\vb\,,
\end{equation}
where
\[
\int\limits_\bV \nabla_\vb\cdot\, S(\vb,y,O_2,M) p\, d\vb=0
\]
for the boundary conditions imposed on $\vb$. For the proliferative operator we have either
\[
\int\limits_\bV\mathcal{P}_1[p]d\vb=\int\limits_\bV \mu(1-s^*)\left(1-M\right)\dfrac{O_2}{1+O_2}p(t,\x,\vb,y)d\vb= \mu(1-s^*)\left(1-M\right)\dfrac{O_2}{1+O_2} m(t,\x,y)=\mathcal{P}_1[m]
\]
or
\begin{align*}
\int\limits_\bV\mathcal{P}_2[p]d\vb&=\int\limits_\bV \mu\left(1-M\right)\dfrac{O_2}{1+O_2}\int\limits_\bY(1-y')\chi(t,\x,y)p(t,\x,\vb,y')dy'd\vb\\[0.3cm]
&=\mu\left(1-M\right)\dfrac{O_2}{1+O_2}\int\limits_\bY(1-y')\chi(t,\x,y)m(t,\x,y')dy'=\mathcal{P}_2[m]\,.
\end{align*}
Thus, the equation for $m(t,\x,y)$ reads
\begin{equation}\label{m_eq1}
\varepsilon^\kappa\dfrac{\partial m}{\partial t} +\varepsilon\nabla_\x \cdot\,m^\vb +\varepsilon^{-\nu}\dfrac{\partial }{\partial y}(G(y,O_2)m)=\varepsilon^\kappa\mathcal{P}_k[m]\,.
\end{equation}
Then, we define the vector $m^\vb$ as
\begin{equation*}
m^\vb:=
\begin{pmatrix}
m_1^\vb\\[0.1cm]
m_2^\vb\\[0.1cm]
...\\[0.1cm]
m_n^\vb
\end{pmatrix}
.
\end{equation*}
Considering again \eqref{p_mes_rescale}, multiplying it by $v_i$, and integrating w.r.t $\vb$ we obtain 
\begin{equation*}
\varepsilon^\kappa\dfrac{\partial m_i^\vb}{\partial t} +\varepsilon \int\limits_\bV v_i\nabla_\x \cdot\,(\vb p)d\vb +\varepsilon^{-\nu}\int\limits_\bV v_i \dfrac{\partial }{\partial y}(G(y,O_2)p)d\vb+\int\limits_\bV v_i \nabla_\vb\cdot\,(S(\vb,y,O_2,M) p)d\vb=\varepsilon^\kappa\int\limits_\bV v_i \mathcal{P}_k[p]d\vb\,,
\end{equation*}
where $m_i^\vb(t,\x,y)$ is the $i$-th component of the vector $m^\vb$. Here,
\begin{equation}
\begin{split}
\int\limits_\bV v_i\nabla_\x \cdot\,(\vb p)d\vb &=\nabla_\x \cdot \int\limits_\bV v_i (\vb p)d\vb=\nabla_\x \cdot\int\limits_\bV (v_i-{v}_i^*)(\vb-{\vb}^*)pd\vb+\nabla_\x \cdot\int\limits_\bV (v_i{\vb}^*+{v}_i^*\vb-{v}_i^*{\vb}^*)pd\vb\\[0.4cm]
&=\nabla_\x\cdot ({\vb}^*m_i^\vb+{v}_i^*m^\vb-{v}_i^*{\vb}^*m)
\end{split}
\end{equation}
for the assumption in Section \ref{Sec_assum}, while

\begin{equation*}
\begin{split}
\int\limits_\bV v_i \nabla_\vb\cdot\,(S(\vb,y,O_2,M) p)d\vb&=\int\limits_\bV v_i \partial_{v_i}(S_ip)d\vb+\sum_{\substack{j=1\\ j\ne i}}^N\,\int\limits_\bV v_i\partial_{v_j}(S_jp)d\vb\\[0.2cm]
&=\int\limits_{\bigcup\limits_{\substack{j=1\\j\ne i}}^N\bV_j}\int\limits_{\bV_i} v_i \partial_{v_i}(S_ip)\,dv_i \,d\tilde{\vb}+\sum_{\substack{j=1\\ j\ne i}}^N\,\,\int\limits_{\bigcup\limits_{\substack{k=1\\k\ne j}}^N\bV_k} v_i\int\limits_{\bV_j}\partial_{v_j}(S_jp)\,dv_j\, d\tilde{\vb}
\end{split}
\end{equation*}
where $d\tilde{\vb}$ is {\it (n-1)}-th components vector such that 
$$\vb=(v_i,\tilde{\vb})\in\bV_i\times \bigcup\limits_{\substack{j=1\\j\ne i}}\bV_j=\bV\,.$$ 
Under the boundary conditions in Section \ref{Sec_assum}, the second term on the right-hand side is equal to 0. Using the chain rule the first term reduces to
\begin{equation*}
\begin{split}
\int\limits_{\bigcup\limits_{\substack{j=1\\j\ne i}}^N\bV_j}\int\limits_{\bV_i} v_i \partial_{v_i}(S_ip)\,dv_i \,d\tilde{\vb}&=\int\limits_{\bigcup\limits_{\substack{j=1\\j\ne i}}^N\bV_j}\Big[v_i\,(S_i p)_{|\partial\bV_i}-\int\limits_{\bV_i}S_ipdv_i\Big]d\tilde{\vb}=-\int\limits_\bV S_ipd\vb\\[0.2cm]
&=- a_1\,y\,\left(\beta\dfrac{(\nabla_\x O_2)_i}{\sqrt{1+|\nabla_\x O_2|^2}}-(1-\beta)\dfrac{(\nabla_\x M)_i}{\sqrt{1+|\nabla_\x M|^2}}\right)m+a_2m_i^\vb\\[0.3cm]
&=-g_i(y,O_2,M)m+a_2m_i^\vb\,.
\end{split}
\end{equation*}
Here, $g_i(y,O_2,M)$ represents the $i$-th component of the vector function $g(y,M,O_2)$ defined in \eqref{micro_g}. Concerning the proliferative operator, we have either
\[
\int\limits_\bV v_i\mathcal{P}_1[p]d\vb=\int\limits_\bV v_i \mu(1-s^*)\left(1-M\right)\dfrac{O_2}{1+O_2}p(t,\x,\vb,y)d\vb= \mu(1-s^*)\left(1-M\right)\dfrac{O_2}{1+O_2} m_i^\vb(t,\x,y)=\mathcal{P}_1[m_i^\vb]
\]
or
\begin{align*}
\int\limits_\bV v_i\mathcal{P}_2[p]d\vb&=\int\limits_\bV v_i\mu\left(1-M\right)\dfrac{O_2}{1+O_2}\int\limits_\bY(1-y')\chi(t,\x,y)p(t,\x,\vb,y')dy'd\vb\\[0.3cm]
&=\mu\left(1-M\right)\dfrac{O_2}{1+O_2}\int\limits_\bY(1-y')\chi(t,\x,y)m_i^\vb(t,\x,y')dy'=\mathcal{P}_2[m_i^\vb]\,.
\end{align*}
Thus, the equation for $m_i^{\vb}$ reads
\begin{equation}\label{miv_eq}
\varepsilon^\kappa\dfrac{\partial m_i^\vb}{\partial t} +\varepsilon \nabla_\x\cdot (\vb^*m_i^\vb+v_i^*m^\vb-\varepsilon v_i^*\vb^*m) +\varepsilon^{-\nu}\dfrac{\partial }{\partial y}(G(y,O_2)m_i^\vb)-(g_i(y,O_2,M)m-a_2m_i^\vb)=\varepsilon^\kappa\mathcal{P}_k[m_i^\vb]\,.
\end{equation}
Therefore, the system for the n+1 variables $(m,m_1^\vb,m_2^\vb,..,m_n^\vb)$ is given by
\begin{equation}\label{sis_m}
\begin{sistem}
\varepsilon^{\kappa+\nu}\dfrac{\partial m}{\partial t}(t,\x,y) +\varepsilon^{1+\nu}\nabla_\x \cdot\,m^\vb(t,\x,y) +\dfrac{\partial }{\partial y}\Big(G(y,O_2)m(t,\x,y)\Big)=\varepsilon^{\kappa+\nu}\mathcal{P}_k[m](t,\x,y)\,,\\[1cm]
\begin{split}
\varepsilon^{\kappa+\nu}\dfrac{\partial m_i^\vb}{\partial t}(t,\x,y) &+\varepsilon^{1+\nu}\nabla_\x\cdot \big[\vb^*m_i^\vb+v_i^*m^\vb-v_i^*\vb^*m\big](t,\x,y) +\dfrac{\partial }{\partial y}\big[G(y,O_2)m_i^\vb(t,\x,y)\big]\\[0.7cm]
&=\varepsilon^{\nu}\big[g_i(y,O_2,M)m(t,\x,y)-a_2m_i^\vb(t,\x,y)\big]+\varepsilon^{\kappa+\nu}\mathcal{P}_k[m_i^\vb](t,\x,y)\quad\qquad\forall i=1...n\,.
\end{split}
\end{sistem}
\end{equation}
We remark that $\nabla_\x\cdot m^\vb=\sum\limits_{j=1}^n\partial_{x_j}m_j^\vb$. We now integrate \eqref{m_eq1} w.r.t $y$. For the assumptions in Section \ref{Sec_assum}, we immediately get
\begin{equation}\label{M_eq}
\varepsilon^\kappa\dfrac{\partial M}{\partial t}(t,\x)+\varepsilon\nabla_\x\cdot M^\vb(t,\x)=\varepsilon^\kappa\int_\bY\mathcal{P}_k[m]\,
\end{equation}
where for the proliferative operator we have either
\[
\int\limits_\bY\mathcal{P}_1[m]dy=\int\limits_\bY \mu(1-s^*)\left(1-M\right)\dfrac{O_2}{1+O_2}m(t,\x,y)dy= \mu(1-s^*)\left(1-M(t,\x)\right)\dfrac{O_2}{1+O_2}M(t,\x)
\]
or
\begin{align*}
\int\limits_\bY\mathcal{P}_2[m]dy&=\int\limits_\bY \mu\left(1-M(t,\x)\right)\dfrac{O_2}{1+O_2}\int\limits_\bY(1-y')\chi(t,\x,y)m(t,\x,y')dy'dy\\[0.3cm]
&=\mu\left(1-M\right)\dfrac{O_2}{1+O_2}\iint\limits_\bY(1-y')\chi(t,\x,y)m(t,\x,y')dy'dy\\[0.3cm]
&=\mu\left(1-M(t,\x)\right)\dfrac{O_2}{1+O_2}\left(M(t,\x)-M^y(t,\x)\right)\,.
\end{align*}
Instead, integrating \eqref{miv_eq} w.r.t. $y$, we obtain 
\begin{equation*}
\begin{split}
&\varepsilon^\kappa\dfrac{\partial M_i^\vb}{\partial t} +\varepsilon\nabla_\x \cdot\left(\vb^*M_i^\vb+v_i^*M^\vb-v_i^*\vb^*M\right) +\varepsilon^{-\nu}\int\limits_\bY \dfrac{\partial }{\partial y}(G(y,O_2)m_i^\vb)dy-\int\limits_\bY (g_i(y,O_2,M)m-a_2m_i^\vb)dy\\[0.3cm]
&=\varepsilon^\kappa\int\limits_\bY\mathcal{P}_k[m_i^\vb]dy\,,
\end{split}
\end{equation*}
where the first integral vanishes for the boundary conditions in Section \ref{Sec_assum}. Concerning the second integral on the left-hand side, we have
\[
\int\limits_\bY (g_i(y,O_2,M)m-a_2m_i^\vb)dy=\tilde{g}_i(O_2,M)M^y-a_2M_i^\vb\,,
\]
where
\begin{equation}\label{gitilde}
    \tilde{g}_i(O_2,M)=a_1\,\left(\beta\dfrac{(\nabla_\x O_2)_i}{\sqrt{1+|\nabla_\x O_2|^2}}-(1-\beta)\dfrac{(\nabla_\x M)_i}{\sqrt{1+|\nabla_\x M|^2}}\right)\,.
\end{equation}
The proliferative operator reads either
\[
\int\limits_\bY\mathcal{P}_1[m_i^\vb]dy=\int\limits_\bY \mu(1-s^*)\left(1-M\right)\dfrac{O_2}{1+O_2}m_i^\vb(t,\x,y)dy= \mu(1-s^*)\left(1-M\right)\dfrac{O_2}{1+O_2} M_i^\vb(t,\x)
\]
or
\begingroup
\allowdisplaybreaks
\begin{align*}
\int\limits_\bY \mathcal{P}_2[m_i^\vb]dy&=\int\limits_\bY \mu\left(1-M\right)\dfrac{O_2}{1+O_2}\int\limits_\bY(1-y')\chi(t,\x,y)m_i^\vb(t,\x,y')dy'dy\\[0.3cm]
&=\mu\left(1-M\right)\dfrac{O_2}{1+O_2}\left(M_i^\vb-\iint\limits_\bY y'\chi(t,\x,y)m_i^\vb(t,\x,y')dy'dy\right)\\[0.3cm]
&=\mu\left(1-M\right)\dfrac{O_2}{1+O_2}\left(M_i^\vb-\int\limits_\bY \int\limits_\bV (v_i-v_i^*)y\,p(t,\x,\vb,y)dyd\vb-v_i^*\int\limits_\bY\int\limits_\bV y\,p(t,\x,\vb,y)dyd\vb\right)\\[0.3cm]
&=\mu\left(1-M\right)\dfrac{O_2}{1+O_2}\Bigg(M_i^\vb-\int\limits_\bY \int\limits_\bV (v_i-v_i^*)(y-y^*)\,p(t,\x,\vb,y)dyd\vb\\[0.3cm]
&\,\,\,\,-y^*\int\limits_\bY\int\limits_\bV(v_i-v_i^*)p(t,\x,\vb,y)dyd\vb-v_i^*\int\limits_\bY\int\limits_\bV y\,p(t,\x,\vb,y)dyd\vb\Bigg)\\[0.3cm]
&=\mu\left(1-M(t,\x)\right)\dfrac{O_2}{1+O_2}\left(M_i^\vb(t,\x)-y^*M_i^\vb(t,\x)-v_i^*M^y(t,\x)+y^*v_i^*M(t,\x)\right)\,.
\end{align*}
\endgroup
Therefore, we obtain
\begin{equation}\label{eq_Miv}
\begin{split}
\varepsilon^\kappa\dfrac{\partial M_i^\vb}{\partial t}(t,\x) &+\varepsilon\nabla_\x \cdot\left[\vb^*M_i^\vb+v_i^*M^\vb- v_i^*\vb^*M\right](t,\x) -\left[\tilde{g}_i(O_2,M)M^y(t,\x)-a_2M_i^\vb(t,\x)\right]=\varepsilon^\kappa\int\limits_\bY\mathcal{P}_k[m_i^\vb]dy\,.
\end{split}
\end{equation}
Finally, multiplying \eqref{m_eq1} by $y$ and integrating w.r.t. $y$ we get 
\begin{equation*}
\varepsilon^\kappa\dfrac{\partial M^y}{\partial t}+\varepsilon \nabla_\x \cdot\int\limits_\bY y M^\vb dy +\varepsilon^{-\nu}\int\limits_\bY\,y\dfrac{\partial }{\partial y}\Big(G(y,O_2)m\Big) dy=\varepsilon^\kappa\int\limits_\bY y\,\mathcal{P}_k[m] (t,\x,y)dy\,,
\end{equation*}
where
\begin{equation*}
\begin{split}
\nabla_\x \cdot\int\limits_\bY y M^\vb dy=\nabla_\x \cdot\int\limits_\bY\int\limits_\bV\,y (\vb p)d\vb dy&=\nabla_\x \cdot\left(\int\limits_\bY\int\limits_\bV (\vb-{\vb}^*)(y-y^*)pd\vb dy +\int\limits_\bY\int\limits_\bV ({\vb}^*y+\vb y^*-{\vb}^*y^*)pd\vb dy\right) \\[0.2cm]
&=\nabla_\x \cdot(\vb^*M^y+y^*M^\vb-\vb^*y^*M)\,,
\end{split}
\end{equation*}
thanks to the assumption in Section \ref{Sec_assum}, while

\begin{align*}
\int\limits_\bY\,y\dfrac{\partial }{\partial y}\Big({G}(y,O_2)m\Big) dy&=\left(y({G}m)_{|\partial\bY}-\int\limits_\bY{G}(y,O_2)mdy\right)=-\int\limits_\bY{G}(y,O_2)mdy=-\int\limits_\bY\left(g_s\dfrac{1}{1+y}{H}(O_2)-\gamma_sy\right)mdy\\[0.2cm]
&=\gamma_s M^y-g_s{H}(O_2) \int\limits_\bY \dfrac{1}{1+y}m dy\,.
\end{align*}
Considering the Taylor expansion of $\dfrac{1}{1+y}$ around $y^*$ we get
\[
\dfrac{1}{1+y}=\dfrac{1}{1+y^*+y-y^*}\approx (1+y^*)^{-1}-(1+y^*)^{-2}(y-y^*)+(1+y^*)^{-3}(y-y^*)^2+O((y-y^*)^2)\,.
\]
Thus, ignoring the higher-order terms of this expansion, we obtain
\begin{equation*}
\begin{split}
\int\limits_\bY \dfrac{1}{1+y}m dy&=\int\limits_\bY \left((1+y^*)^{-1}-(1+y^*)^{-2}(y-y^*)+(1+y^*)^{-3}(y-y^*)^2\right)m\,dy\\[0.4cm]
&=(1+y^*)^{-1} M- (1+y^*)^{-2} M^y+y^*(1+y^*)^{-2}M\,,
\end{split}
\end{equation*}
i.e.,
\begin{equation*}
\begin{split}
\int\limits_\bY\,y\dfrac{\partial }{\partial y}\Big({G}(y,O_2)p\Big) dy&=\gamma_sM^y-g_s\dfrac{{H}(O_2)}{1+y^*} \left(M-\dfrac{1}{1+y^*}(M^y-y^*M)\right)\\[0.4cm]
&=-g_s\dfrac{{H}(O_2)}{(1+y^*)^2}(1+2y^*)M+\left(\gamma_s+g_s\dfrac{{H}(O_2)}{(1+y^*)^2}\right)M^y\,.
\end{split}
\end{equation*}
The proliferative operator in this case reads either
\[
\int\limits_\bY y\mathcal{P}_1[m]dy=\int\limits_\bY \mu(1-s^*)\left(1-M\right)\dfrac{O_2}{1+O_2}ym(t,\x,y)dy= \mu(1-s^*)\left(1-M(t,\x)\right)\dfrac{O_2}{1+O_2} M^y(t,\x)
\]
or
\begin{align*}
\int\limits_\bY y \mathcal{P}_2[m]dy&=\int\limits_\bY \mu\left(1-M\right)\dfrac{O_2}{1+O_2}\int\limits_\bY y(1-y')\chi(t,\x,y)m(t,\x,y')dy'dy\\[0.3cm]
&=\mu\left(1-M\right)\dfrac{O_2}{1+O_2}\left(\int\limits_\bY y\chi(t,\x,y)dy\int\limits_\bY (1-y')m(t,\x,y')dy'\right)\\[0.3cm]
&=\mu\left(1-M\right)\dfrac{O_2}{1+O_2}\left(M-M^y\right)\left(\int\limits_\bY (y-y^*)\chi(t,\x,y)dy+y^*\int\limits_\bY\chi(t,\x,y)dy\right)\\[0.3cm]
&=\mu\left(1-M(t,\x)\right)\dfrac{O_2}{1+O_2}y^*\left[M(t,\x)-M^y(t,\x)\right]\,.
\end{align*}
Thus, the resulting equation for $M^y$ is given by
\begin{equation}
\begin{split}
&\varepsilon^\kappa\dfrac{\partial M^y}{\partial t}(t,\x) +\varepsilon\nabla_\x \cdot\left[\vb^*M^y+y^*M^\vb-\vb^*y^*M\right](t,\x) -\varepsilon^{-\nu}g_s\dfrac{{H}(O_2)}{(1+y^*)^2}(1+2y^*)M(t,\x)\\[0.3cm]
&+\varepsilon^{-\nu}\left(\gamma_s+g_s\dfrac{{H}(O_2)}{(1+y^*)^2}\right)M^y(t,\x)=\varepsilon^\kappa \int\limits_\bY y\mathcal{P}_k[m](t,\x,y)dy\,.
\end{split}
\end{equation}
Therefore, the system for $M(t,\x)$, $M^\vb(t,\x)$, and $M^y(t,\x)$ reads
\begin{equation}\label{sis_M}
\begin{sistem}
\varepsilon^{\kappa-1}\dfrac{\partial M}{\partial t}(t,\x)+\nabla_\x\cdot M^\vb(t,\x)=\varepsilon^{\kappa-1}\displaystyle\int\limits_\bY\mathcal{P}_k[m](t,\x,y)dy\,,\\[1cm]
\begin{split}
&\varepsilon^{\kappa}\dfrac{\partial M_i^\vb}{\partial t}(t,\x)+\varepsilon\nabla_\x \cdot\left[\vb^*M_i^\vb+v_i^*M^\vb- v_i^*\vb^*M\right](t,\x) -(\tilde{g}_i(O_2,M)M^y(t,\x)-a_2M_i^\vb(t,\x))\\[0.3cm]
&=\varepsilon^{\kappa}\displaystyle\int\limits_\bY\mathcal{P}_k[m_i^\vb](t,\x,y)dy\,,\qquad\qquad\qquad\qquad\forall i=1...n\,,\end{split}
\\[1.2cm]
\begin{split}
&\varepsilon^{\kappa+\nu}\dfrac{\partial M^y}{\partial t}(t,\x)  +\varepsilon^{1+\nu} \nabla_\x \cdot\left[\vb^*M^y+y^*M^\vb-\vb^*y^*M\right](t,\x) -g_s\dfrac{{H}(O_2)}{(1+y^*)^2}(1+2y^*)M(t,\x)\\[0.3cm]
&+\left( \gamma_s+g_s\dfrac{{H}(O_2)}{(1+y^*)^2}\right)M^y(t,\x)=\varepsilon^{\kappa+\nu}\int\limits_\bY y\mathcal{P}_k[m](t,\x,y)dy\,.
\end{split}
\end{sistem}
\end{equation}
Now, we consider the expansion of the previously introduced moments in the form
\begin{align*}
&M=M_0+\varepsilon M_1+O(\varepsilon^2)\,,\\  
&M^\vb=M_0^\vb+\varepsilon M_1^\vb+O(\varepsilon^2)\,,\\
&M^y=M^y_0+\varepsilon M^y_1+O(\varepsilon^2)\,,
\end{align*}
such that $M_0=\lim\limits_{\varepsilon\to 0}M$. Following the well-established literature for the modeling of directed cell migration in response to environmental cues \cite{hillen2013transport,bellomo2010complexity,bellomo2007multicellular,filbet2005derivation,conte2022multi}, we consider a hyperbolic limit of the moment system \eqref{sis_M} by choosing $\kappa=1$. Passing formally to the limit $\varepsilon\to 0$ in \eqref{sis_M}, from the first equation we get
\begin{equation}\label{M0_eq}
\dfrac{\partial M_0}{\partial t}(t,\x)+\nabla_\x\cdot M_0^\vb(t,\x)=\displaystyle\int\limits_\bY\mathcal{P}_k[m_0](t,\x,y)dy\,,\\    
\end{equation}
where $m_0$ is the zero-order term in the expansion of the moment $m(t,\x,y)$. Then, from the equation for $M_i^\vb$ in \eqref{sis_M}, we get
\begin{equation}
 \tilde{g}_i(O_2,M_0)M_0^y-a_2M_{i,0}^\vb=0\,\,\,\Longleftrightarrow\,\,\, M_{i,0}^\vb=\dfrac{\tilde{g}_i(O_2,M_0)}{a_2}M_0^y  
\end{equation}
that, considering the vector $M_0^\vb$ with $i$-th component $M_{i,0}^\vb$, means 
\begin{equation}\label{MV_eq}
    M_0^\vb=\dfrac{{\bf \tilde{g}}(O_2,M_0)}{a_2}M_0^y  
\end{equation}
with 
\begin{equation*}
    {\bf \tilde{g}}(O_2,M_0)=a_1\,\left(\beta\dfrac{\nabla_\x O_2}{\sqrt{1+|\nabla_\x O_2|^2}}-(1-\beta)\dfrac{\nabla_\x M_0}{\sqrt{1+|\nabla_\x M_0|^2}}\right)\,.
\end{equation*}
Finally, from the last equation in \eqref{sis_M}, we obtain
\begin{equation}\label{MY_eq}
-g_s\dfrac{{H}(O_2)}{(1+y^*)^2}(1+2y^*)M_0+\left(\gamma_s+g_s\dfrac{{H}(O_2)}{(1+y^*)^2}\right)M_0^y=0 ,\,\,\Longleftrightarrow\,\,\,M_0^y=\dfrac{g_s{H}(O_2)(1+2y^*)}{\gamma_s(1+y^*)^2+g_s{H}(O_2)}M_0\,.
\end{equation}
Thus, plugging \eqref{MY_eq} into \eqref{MV_eq}, and the latter in \eqref{M0_eq}, we obtain the following equation for the evolution of the macroscopic cell density $M_0$
\begin{equation}\label{Macro_eq}
\dfrac{\partial M_0}{\partial t}(t,\x)+\nabla_\x\cdot \left[ F(y^*,O_2(\x))M_0(t,\x)\left(\beta\dfrac{\nabla_\x O_2(\x)}{\sqrt{1+|\nabla_\x O_2(\x)|^2}}-(1-\beta)\dfrac{\nabla_\x M_0(t,\x)}{\sqrt{1+|\nabla_\x M_0(t,\x)|^2}}\right) \right]=\mathbb{P}_k[M_0](t,\x)
\end{equation}
where we set the velocity field  (or chemotactic sensitivity)
\begin{equation}\label{F}
    F(y^*,O_2(\x)):=\dfrac{a_1}{a_2}\dfrac{g_s{H}(O_2(\x))(1+2y^*)}{\gamma_s(1+y^*)^2+g_s{H}(O_2(\x))}\,,
\end{equation}
while the proliferative operator is given by either
\begin{equation}\label{Macro_prol_1}
    \mathbb{P}_1[M_0](t,\x)=\int\limits_\bY\mathcal{P}_1[m_0](t,\x,y)dy=\mu(1-s^*)\left(1-M_0(t,\x)\right)\dfrac{O_2(\x)}{1+O_2(\x)}M_0(t,\x)
\end{equation}
or
\begin{equation}\label{Macro_prol_2}
    \mathbb{P}_2[M_0](t,\x)=\int\limits_\bY\mathcal{P}_2[m_0](t,\x,y)dy=\mu\left(1-M_0(t,\x)\right)\dfrac{O_2(\x)}{1+O_2(\x)}\left(1-\dfrac{a_2}{a_1}F(y^*,O_2(\x))\right)M_0(t,\x)\,.
\end{equation}
Since we are interested in the impact of the Snail protein expression of a cell on the overall macroscopic evolution of the population, in the numerical experiments described in Section \ref{Sec_numsim} we also take into account $M_0^y$, which is the average expression of the Snail protein in the cell population. Discarding the subscripts $_0$, the resulting macroscopic system reads
\begin{equation}\label{sim_M_ups}
\begin{sistem}
\dfrac{\partial M}{\partial t}(t,\x)+\nabla_\x\cdot \left[ F(y^*,O_2(\x))M(t,\x)\left(\beta\dfrac{\nabla_\x O_2(\x)}{\sqrt{1+|\nabla_\x O_2(\x)|^2}}-(1-\beta)\dfrac{\nabla_\x M(t,\x)}{\sqrt{1+|\nabla_\x M(t,\x)|^2}}\right) \right]=\mathbb{P}_k[M](t,\x)\,,\\[0.8cm]
M^y(t,\x)=\dfrac{a_2}{a_1}F(y^*,O_2(\x))M(t,\x)\,.
\end{sistem}
\end{equation}

\section{Numerical experiments}\label{Sec_numsim}
We perform 2D numerical simulations of the resulting macroscopic system \eqref{sim_M_ups} to \textit{in-silico} analyze different scenarios of tumor progression under varied oxygen conditions and Snail expressions. Numerical simulations are conducted using a self-developed code in Python 3.10.12, whose details are provided in Section \ref{Sec_numMet}. The parameter values used for the simulations are reported in Table \ref{parameter_const}. Sections \ref{Sec_numTest1}, \ref{Sec_numTest2}, \ref{Sec_numTest3}, and \ref{Sec_numTest4} present the results of the four numerical experiments.
\hspace{-2cm}{\begin{itemize}
    \item[]
     {\bf Experiment 1 - }In Section \ref{Sec_numTest1}, we analyze the differences in cell migration and Snail distribution over time between a chemotactic--driven scenario and an anti-crowding--driven scenario by varying the value of the weighting parameter $\beta$ and, thus, its impact on cell motion.
    
    \item[]
    {\bf Experiment 2 - }In Section \ref{Sec_numTest2}, we investigate the impact of Snail expression on cell proliferation, and we compare the model's evolution when considering the two different expressions of the proliferative operator given in \eqref{Macro_prol_1} and \eqref{Macro_prol_2}.
    
    \item[] 
   {\bf Experiment 3 - } In Section \ref{Sec_numTest3}, we show the model's capabilities to quantitatively replicate experimental results from two different studies on human cancer cells. Firstly, we consider the findings in \cite{lundgren2009hypoxia} regarding the effect of Snail over-expression or knockdown on the migration capability of human breast cancer cells. Secondly, we refer to the comparison presented in \cite{yu2013notch1}, where the effects of hypoxia and a combination of hypoxia and Snail knockdown on the motility of human hepatocarcinoma cells are studied.
    
    \item[] 
    {\bf Experiment 4 - }In Section \ref{Sec_numTest4}, we analyze the hypoxia-driven spatial distribution of Snail expression within a tumor and we show its consistency with the experimental results shown in \cite{lundgren2009hypoxia}.
\end{itemize}}

\begin{table} [!h]
\begin{center}
   \begin{tabular}{|c|c|c|c|}
   \toprule  
   \rule{0pt}{1.5ex}Parameter & Description & Value (unit) & Source \\
  \midrule
   \rule{0pt}{3ex}$g_s$ & basal Snail transcription rate & 1.5 (molecules $\cdot$ min$^{-1}$)& \cite{lu2013microrna}\\[0.5ex]
\rule{0pt}{3ex}$\gamma_s$ & basal Snail degradation rate & $0.0021$ (min$^{-1}$)&\cite{lu2013tristability}\\[0.5ex]
\rule{0pt}{3ex}$a_1/a_2$ & scaling velocity parameter & 0.1 (mm $\cdot$ min$^{-1}$)  &This work\\[0.5ex]
\rule{0pt}{3ex}$\beta$ & weighting parameter for tactic contribution &varying in [0-1] &This work\\[0.5ex]
\rule{0pt}{3ex}$\mu$ &tumor proliferation rate & [6-9]$\cdot$ 10$^{-4}$ (min$^{-1}$)&\cite{weber2014quantifying,conte2021mathematical}\\[0.5ex]
\bottomrule
    \end{tabular}
\end{center}
\caption{{\bf Model parameters.} The table provides the dimensional values for the parameters involved in setting \eqref{sim_M_ups} and used in the numerical experiments.}
 \label{parameter_const}
\end{table}

\subsection{Numerical method}\label{Sec_numMet}
To perform numerical simulations of the model, we adapt the numerical method presented in \cite{carrillo2015finite} to our problem structure. In detail, we rewrite the first equation of \eqref{sim_M_ups} as
\begin{equation}\label{eq_init_num}
\dfrac{\partial M}{\partial t}(t,\x)= - T(y^*,O_2,M)(t,\x) +\mathbb{P}_k[M](t,\x)
\end{equation}
where
\begin{equation}
\T(y^*,O_2,M)(t,\x) = \nabla_\x\cdot \left[ F(y^*,O_2(\x))M(t,\x)\left(\beta\dfrac{\nabla_\x O_2(\x)}{\sqrt{1+|\nabla_\x O_2(\x)|^2}}-(1-\beta)\dfrac{\nabla_\x M(t,\x)}{\sqrt{1+|\nabla_\x M(t,\x)|^2}}\right) \right]
\end{equation}
rules the movement of cells, while $\mathbb{P}_k[M](t,\x)$ is the proliferation term defined in \eqref{Macro_prol_1} ($k=1$) and \eqref{Macro_prol_2} (${k=2}$). Setting $\x=(x_1,x_2)$, we consider the geometric domain ${\Omega = [x_{1,min}, x_{1,max}]\times[x_{2,min},x_{2,max}]\subseteq \mathbb{R}^2}$, where we introduce a uniform Cartesian mesh consisting of elements
$C_{j,l} := [x_{1,j-\frac{1}{2}}, x_{1,j+\frac{1}{2}}] \times [x_{2,l-\frac{1}{2}}, x_{2,l+\frac{1}{2}}]$, for ${j=0,\dots, N_{x_1}}$ and for ${l=0,\dots, N_{x_2}}$, of size $\Delta x_1   \times\Delta x_2$. We adopt a splitting method, accounting first for the conservative part $\T(y^*,O_2,M)$ and, then, for the reaction term $\mathbb{P}_k[M]$. Precisely, defining 

$$M_{j,l}(t)= \frac{1}{\Delta x_1\Delta x_2} \int\limits_{C_{j,l}} M(t,x_1,x_2) d\mathbf{x}\,,$$
for the conservative part we adopt the general semi-discrete finite-volume scheme given by

$$T_{j+\frac{1}{2},l} = u^+_{j+\frac{1}{2},l}M^E_{j,l} + u^-_{j+\frac{1}{2},l}M^W_{j+1,l}$$
$$T_{j,l+\frac{1}{2}} = v^+_{j,l+\frac{1}{2}}M^N_{j,l} + v^-_{j,l+\frac{1}{2}}M^S_{j,l+1}\,.$$
Here:
\begin{itemize}
    \item $(\cdot)^+$ and $(\cdot)^-$ indicate the positive and negative part of their arguments, respectively, i.e., ${(\cdot)^+= \max\{0,\cdot\}}$ and ${(\cdot)^-= \min\{0,\cdot\}}$;
    
    \item the apices E, W, N, S indicate East, West, North, and South and correspond to the evaluation of the piecewise linear reconstruction using the following first-order truncation of Taylor expansion
    \[
    \tilde{M}(x_1,x_2) = M_{j,l}+ (\partial_{x_1} M)_{j,l}(x_1-x_{1,j}) + (\partial_{x_2} M)_{j,l}(x_2-x_{2,l}), \quad (x_1,x_2)\in C_{j,l}
    \]
    at the element interfaces $(x_{1,j+\frac{1}{2}},x_{2,l})$, $(x_{1,j-\frac{1}{2}},x_{2,l})$, $(x_{1,j},x_{2,l+\frac{1}{2}})$,
    $(x_{1,j},x_{2,l-\frac{1}{2}})$, respectively;
    
    \item defining
    \[
    U_{j,l} = F(y^*(x_{1,j},x_{2,l}),O_2(x_{1,j},x_{2,l}))\left(\beta\dfrac{\nabla_\x O_2(x_{1,j},x_{2,l})}{\sqrt{1+|\nabla_\x O_2(x_{1,j},x_{2,l})|^2}}-(1-\beta)\dfrac{\nabla_\x M(t,x_{1,j},x_{2,l})}{\sqrt{1+|\nabla_\x M(t,x_{1,j},x_{2,l})|^2}}\right)\,,
    \]
    
    then $u := U_{x_1}$ and $v := U_{x_2}$ are the components of $U$ along the horizontal and vertical axis respectively.
\end{itemize}
Note that the derivatives in the middle points are evaluated as

$$(\partial_{x_1} M)_{{j+\frac{1}{2}},l} = \frac{M_{j+1,l}-M_{j,l}}{\Delta x_1}\,, \qquad {(\partial_{x_2}} M)_{j,l+\frac{1}{2}} = \frac{M_{j,l+1}-M_{j,l}}{\Delta x_2}\,,$$
while the derivatives in the nodes are initially evaluated as

$$(\partial_{x_1} M)_{j,l} = \frac{M_{j+1,l}-M_{j-1,l}}{2\Delta x_1}, \qquad
(\partial_{x_2} M)_{j,l} = \frac{M_{j,l+1}-M_{j,l-1}}{2\Delta x_2}$$
and then corrected using a generalized minmod limiter to preserve the positivity of the linear reconstruction $\tilde{M}$ (for further details see \cite{carrillo2015finite}). 
For the time discretization, we use the forward Euler method. We denote with apex $h$ the discretized time step, i.e., 
\[
t^h = t_0 +\sum\limits_{i=1}^{h-1} {\Delta t}_i\,.
\]
To optimize the performances, we use adaptive time steps obtained by imposing the positivity-preserving CFL 
\[
\Delta t_h \leq \Delta \mathcal{T}_h:=\min \left\{\dfrac{\Delta x_1}{4\text{a}},\dfrac{\Delta x_2}{4\text{b}} \right\}
\]
where $\text{a}=\underset{j,l}{\max}\left(|u^{h}_{j+\frac{1}{2},l}|\right)$
and $\text{b}=\underset{j,l}{\max}\left(|v^{h}_{j,l+\frac{1}{2}}|\right)$. Therefore, starting from the discretized initial condition $M_{j,l}^0$ provided for each $j=0,\dots, N_{x_1}$ and for $l=0,\dots,N_{x_2}$, given that $M_{j,l}^h$ is the numerical approximation of $ M_{j,l}(t^h)$, the numerical scheme reads
\begin{equation}\label{num_scheeme}
\begin{sistem}
  M^{h+\frac{1}{2}}_{j,l} =
M^{h}_{j,l}-\dfrac{\Delta t_h}{\Delta x_1} \left(T^{h}_{j+\frac{1}{2},l} - T^{h}_{j-\frac{1}{2},l}\right) - \dfrac{\Delta t_h}{\Delta x_2} \left(T^{h}_{j,l+\frac{1}{2}} - T^{h}_{j,l-\frac{1}{2}}\right) \\[0.5cm]
M^{h+1}_{j,l}  = M^{h+\frac{1}{2}}_{j,l} + \Delta t_h (\mathbb{P}_k)^{h+\frac{1}{2}}_{j,l}
\end{sistem}
\end{equation}
for $h=1,\dots,N_h$. In the proposed experiments, we set the spatial domain $\Omega =  [0,50] \times [0,50] \,\text{mm}^2$ and we consider the time $t \in [0,T]$, with $T>0$. Dealing with a limited domain, we set no entry flux boundary conditions.

\subsection{Experiment 1: chemotactic or anti-crowding -driven motion}\label{Sec_numTest1}
In this first experiment, we compare cell behaviors in two different scenarios. We consider a first scenario in which cell movement is mainly characterized by a chemotactic attraction toward increasing oxygen concentrations, strongly reducing the impact of the natural cell anti-crowding mechanism. Instead, in a second scenario, we emphasize the role of anti-crowding dynamics, which helps cells to avoid highly dense regions.

\begin{figure}[!h]
    \centering
    \includegraphics[width=0.8\linewidth]{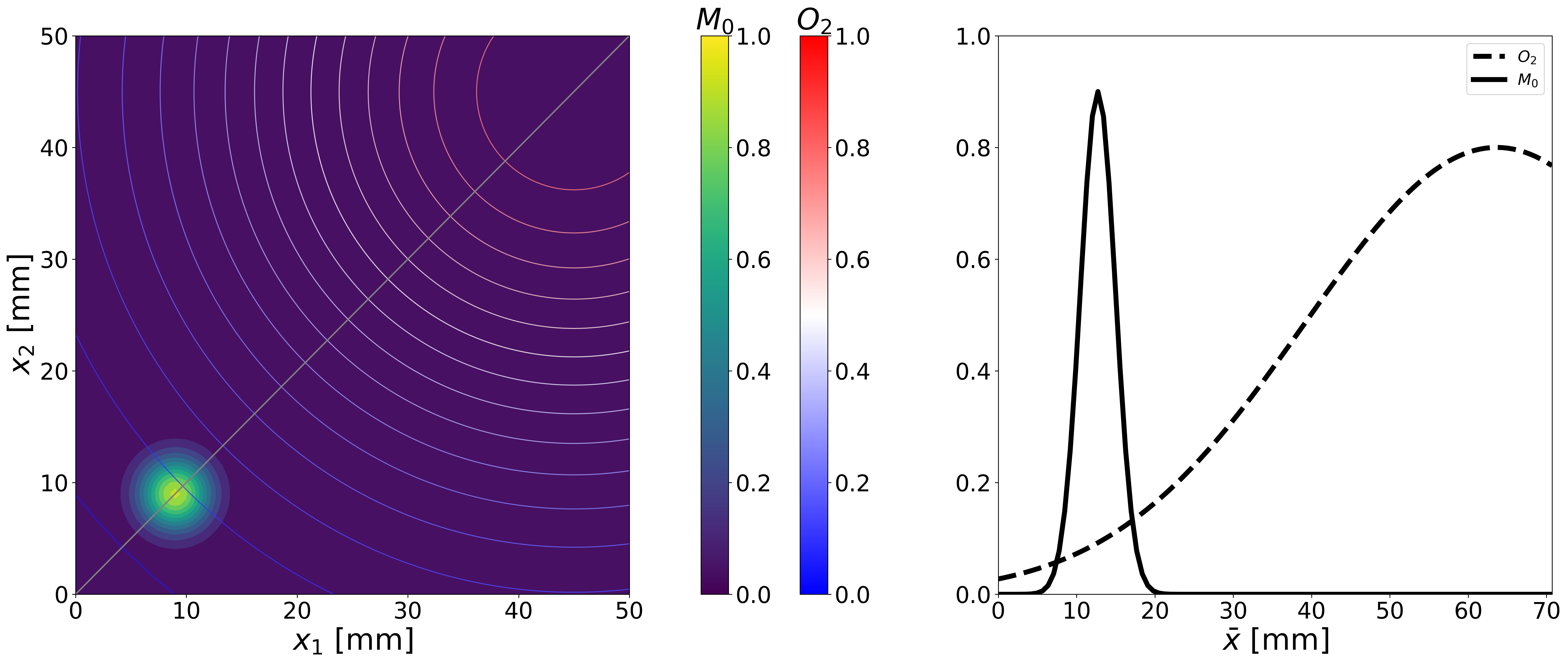}
    \caption{{\bf Experiment 1: initial conditions.} Left: initial Gaussian distribution of the tumor cells $M^0$, centered in $\mathbf{x}_M = [9,9]\,\text{mm}$ in the domain $\Omega=[0,50]\times[0,50]\,\text{mm}^2$, together with the level plot for the fixed Gaussian distribution of oxygen $O_2(\x)$, centered in $\mathbf{x}_O = [45,45]\,\text{mm}$. Right: 1D profiles of tumor (continuous line) and oxygen (dashed line) distributions along the bisecting line (light gray line in the 2D plot) of the domain $\Omega$. $\bar{x}$ indicates the spatial position along this bisecting line.}
    \label{fig:fig1_ic}
\end{figure}

Considering the macroscopic system \eqref{sim_M_ups}, we recall that the parameter $\beta\in[0,1]$ impacts cell motility by weighting the influence of oxygen and cell density gradients on the direction of tumor cell migration. Higher values of $\beta$ imply a stronger impact of chemotaxis compared to anti-crowding dynamics. We choose $\beta=0.98$ to emphasize the role of chemotactic movement toward increasing oxygen concentrations, while $\beta=0.8$ to account for a stronger effect of the anti-crowding mechanism. In this experiment, we consider a fixed oxygen source located in the top-right corner of the domain $\Omega$ and whose expression is given by 

\begin{equation}\label{eq_inO}
\displaystyle{O_2(\mathbf{x}) = I_{O_2}\,e^{-\dfrac{(\mathbf{x}-\mathbf{x}_{O_2})^2}{\theta^2_{O_2}}}}
\end{equation}
with $I_{O_2}=0.8$, $\mathbf{x}_{O_2} = [45, 45]\,\text{mm}$, and $\theta_{O_2}=34\, \text{mm}$. 
For the tumor cells, we assume that the initial tumor mass $M(t=0,\x):=M^0(\x)$ is located in the opposite (bottom-left) corner of the domain $\Omega$ and it is defined as

\begin{equation}\label{eq_inM}
\displaystyle{M^0(\mathbf{x}) = I_M\,e^{-\dfrac{(\mathbf{x}-\mathbf{x}_M)^2}{\theta^2_M}}}
\end{equation}
with $I_M=0.9$, $\mathbf{x}_M = [9, 9]\,\text{mm}$, and $\theta_M=3\, \text{mm}$. The initial configuration of this setting is shown in Figure \ref{fig:fig1_ic}.

\begin{figure}[!h]
    \centering
    \includegraphics[width=0.6\linewidth]{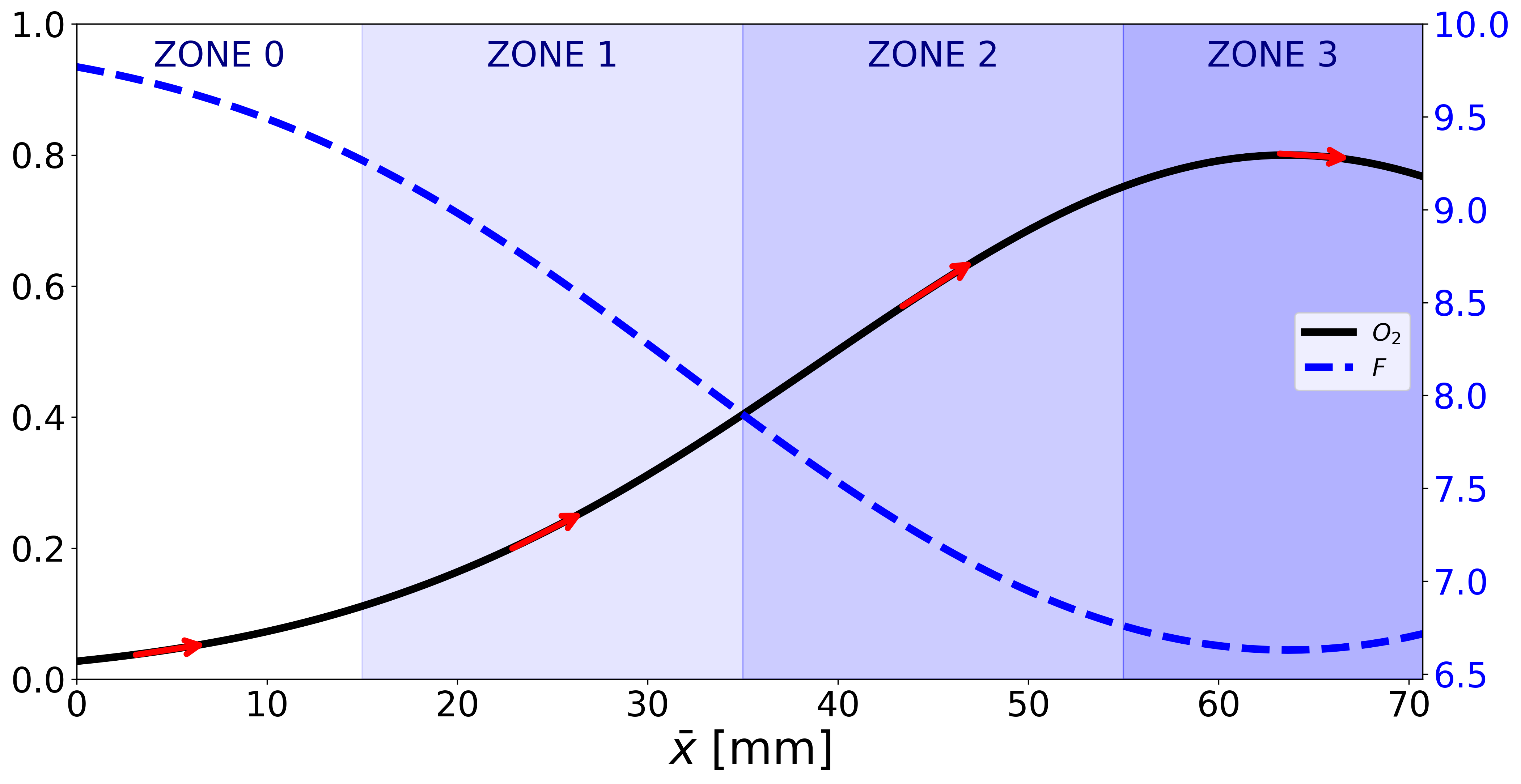}
    \caption{{\bf Experiment 1: spatial distribution of the environmental signals triggering cell motility.} Graphical representation of the 1D section defined by the bisecting line of the domain $\Omega$ and its subdivision into four different areas, depending on the combination of high/low motility, random/directed motion, and low/high proliferation capability of the tumor cells. Profiles of the oxygen distribution (black continuous line) and tactic sensitivity $F(y^{*},O_2(\bar{x}))$ (blue dashed line) are shown, together with the oxygen gradient direction (red arrow). $\bar{x}$ indicates the spatial position along this bisecting line.}
     \label{fig:fig1_phases}
\end{figure}

\noindent To better characterize the dynamics we divide the domain into four different zones depending on the relative oxygen conditions, which trigger cell motility and proliferation. Figure \ref{fig:fig1_phases} and the related Table \ref{table_zone} summarize the combination of the environmental signals in each of these areas along the bisecting line of the domain.

\begin{figure}[!h]
    \centering
    \includegraphics[width=0.7\linewidth]{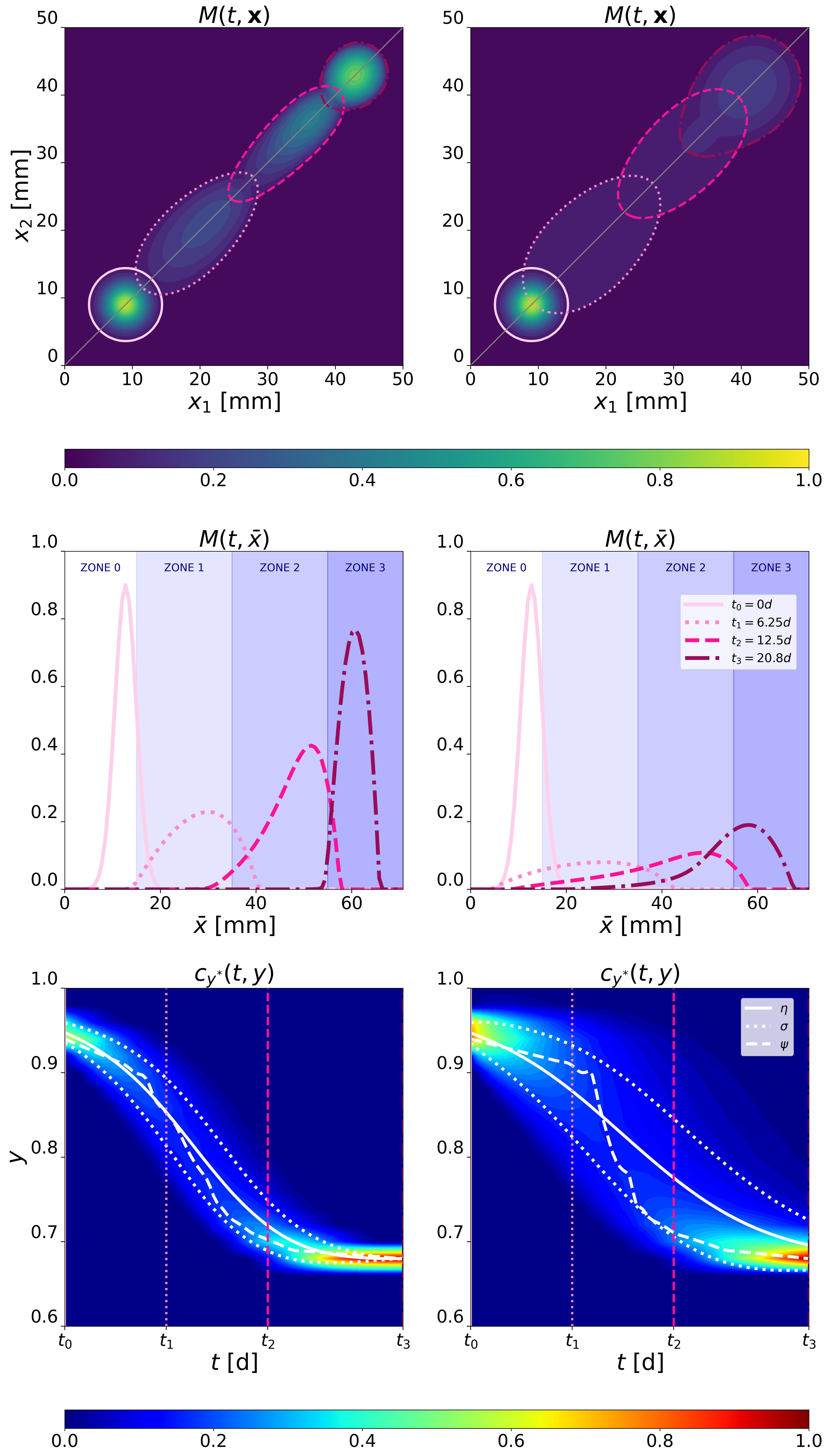}
    \caption{{\bf Experiment 1: chemotactic- or anti-crowding-driven motion.} Evolution of model \eqref{sim_M_ups} in a chemotactic-driven ($\beta=0.98$, left column) or anti-crowding-driven ($\beta=0.8$, right column) scenario. Top row: evolution of the tumor mass in the domain $\Omega=[0,50]\times[0,50]\,\text{mm}^2$ at four different time steps, i.e., initial time $t=0$d (continuous light pink line), and progression at $t=6.25$d (dot pink line), $t=12.5$d (dashed dark pink line), and $t=20.8$d (dot-dashed purple line). Middle row: 1D profiles of the tumor mass evolution along the bisecting line (light gray line in the 2D plot) of the domain $\Omega$ at the same time steps used in the top row. $\bar{x}$ indicates the spatial position along this bisecting line. Bottom row: evolution in the $(t,y)$-space of the distribution $c_{y^*}(t,y)$, together with its mean $\eta(t)$, standard deviation $\sigma(t)$, and mode $\psi(t)$. References to the four selected time steps are repeated in each graph using consistent color and line styles. Parameter values are set as reported in Table \ref{parameter_const}.}
    \label{fig:fig1}
\end{figure}

\begin{table}[!h]
\begin{center}
   \begin{tabular}{|c|c|c|c|c|c|c|}
   \toprule  
   \rule{0pt}{1.5ex} Zone & $O_2$ & $\nabla O_2$ & $F$ & motility & direction & proliferation\\
  \midrule
   \rule{0pt}{3ex}  0 &  low & low & high & high  & random & low \\[0.5ex]
   \rule{0pt}{3ex}  1 &  low & high & high & high  & directed & low \\[0.5ex]
   \rule{0pt}{3ex}  2 &  high & high & low  & low &  directed & high \\[0.5ex]
   \rule{0pt}{3ex}  3 &  high & low & low & low &  random & high \\[0.5ex]
   \bottomrule
    \end{tabular}
\end{center}
\caption{{\bf Summary of the environmental signals.} The table reports and summarizes the information received by the cells in the four identified areas shown in Figure \ref{fig:fig1_phases}.}
 \label{table_zone}
\end{table} 
\noindent The results of this first experiment are illustrated in Figure \ref{fig:fig1}. Precisely, the left column refers to the simulations obtained for $\beta=0.98$ (chemotactic--driven scenario), while the right column to $\beta=0.8$ (anti-crowding--driven scenario). 
\noindent The simulations track the evolution of the tumor mass over a time range of approximately $T=21$ days. In the first row of Figure \ref{fig:fig1}, we illustrate the progression of the tumor mass in the domain $\Omega$ at four different temporal steps. The contour plot illustrates the density map, while the contour lines represent the tumor's edge (defined by a density threshold corresponding to $10\%$ of the carrying capacity). To better visualize the differences in the tumor evolution in the two scenarios we consider a 1D section of the domain $\Omega$ illustrating the tumor profile along the bisecting line. Moreover, to account for the epigenetic trait information, for every value $y \in \bY$ and every time $t$, we evaluate the fraction of tumor mass having an equilibrium Snail expression $y^*$ lower than $y$, i.e.,
\[
C_{y^*}(t, y) = \int\limits_{D_y}\bar{M}(t,\x) d\x
\]
where $D_{y}=\left\{\x\in\Omega:y^*(\x)\in(-\infty,y)\right\}$ and $\bar{M}(t,\x)$ is the normalized tumor density distribution. Moreover, we introduce the quantity
\[
c_{y^*}(t, y)=\partial_y C_{y^*}(t,y)\,,
\]
which indicates the fraction of mass that, at a given time $t$, has a certain Snail expression $y \in \bY$. For this quantity, we evaluate mean $\eta(t)$, mode $\psi(t)$, and standard deviation $\sigma(t)$ as
\begin{equation*}
\begin{split}
&\eta(t) = \int\limits_{\bY} y c_{y^*}(t,y) dy \,,\\[0.3cm]
&\psi(t) = \argmaxA\limits_{y\in\bY} \left(c_{y^*}(t,y)\right)\,, \\[0.3cm] 
&\sigma(t) = \sqrt{\int\limits_{\bY} y^2 c_{y^*}(t,y) dy - \eta(t)^2}\,.
\end{split}    
\end{equation*}
 Combining the information about the environmental signals (Figure \ref{fig:fig1_phases}) with the data concerning the expression of Snail (third row of Figure \ref{fig:fig1}), we can better grasp the features characterizing the two depicted scenarios. Depending on $\beta$, cell behaviors towards more oxygenated regions show evident differences. Low oxygen concentration corresponds to high Snail expression and they collectively contribute both to a high tactic sensitivity $F(y^{*},O)$ and a low proliferation rate. Initially, as the mass is situated in an area characterized by low oxygenation, this oxygen deprivation triggers cell motility. As the mass progresses towards the upper right corner (temporal step $t_1$), the low oxygen levels still induce a more motile than proliferative cell phenotype, enforced by a high value of the mean $\eta(t)$ (depicted in the third row of Figure \ref{fig:fig1}). By the time $t_1$, the mass has spread enough to develop a smoother profile, resulting in a reduced density gradient in favor of stronger chemotactic motion due to the oxygen gradient. This is particularly evident for $\beta=0.98$. Cells are accelerated and move compactly towards the oxygen source, and this is evidenced in both columns of Figure \ref{fig:fig1} by the reduction in mass width orthogonal to the chemotactic gradient between times $t_1$ and $t_2$, confirming cell convergence towards the oxygen distribution. As the cells reach areas closer to the source (temporal step $t_2$), a still strong oxygen gradient is countered by increased oxygenation levels, and, thus, inhibition of motility in favor of an enhanced proliferation rate. This results also in a lower level of $\eta(t)$ (third row of Figure \ref{fig:fig1}). This combination results in masses developing a distinct tail to the left, with some cells remaining outside the region of orderly motility due to lower oxygen levels. Many motile cells originating from less oxygenated areas "push" against a slowing front where cells are less motile, but contribute to the increase of the density due to their proliferative capability. At the final time $t_3$, the mass has reached a region characterized by high oxygenation levels with an almost negligible oxygen gradient. At this stage, ordered motile dynamics become nearly absent and the mean $\eta(t)$ has reached a lower stable value. This is particularly evident in the scenario with $\beta=0.98$, where the mass tends to regain a more radial symmetry and slightly expands under the influence of density pressure driven by proliferative dynamics and anti-crowding.\\
\indent From a qualitative viewpoint, in both scenarios the dynamics are initially characterized by a diffusing mass that expands orthogonally to the direction of chemotactic motion (more evident for $\beta = 0.8$) and then by a direct movement towards the oxygen source. Moreover, in both cases, the presence of flux-limited operators in the drift term determines steep and well-defined invasion fronts, reducing the typical artificial tails characterizing linear diffusion and, thus, its excessive influence on cell spread. However, comparing the experiments conducted for the two scenarios makes it evident how conditions favoring anti-crowding dynamics can lead to significant changes in tumor shape even for small variations in $\beta$. In fact, when anti-crowding drives the dynamics, cells tend to move toward the location of the oxygen source with a large spread in the domain, with respect to the chemotactic-driven scenario, which shows cells compactly migrating towards more oxygenated regions. Looking at the tumor profiles at time $t_1$ and $t_2$ (second row in Figure \ref{fig:fig1}), the differences in height and the size of the mass support are evident. This is reflected also in the evolution of mean, mode, and standard deviation of $c_{y^*}(y,t)$. In fact, for higher values of $\beta$, mean $\eta$ and mode $\psi$ show a similar trend over time. However, the lower $\beta$, the greater the differences in their evolution. This is because the larger spread of the tumor mass and the slower cell movement in the domain keep the value of the mean higher for a longer time. Moreover, this determines the wider variety of values covered by the distribution and, thus, a larger standard deviation $\sigma$.

\subsection{Experiment 2: impact of Snail expression on cell proliferation}\label{Sec_numTest2}
In the second experiment, we focus on the proliferative dynamics characterizing tumor cells. Here, we assume that chemotaxis drives cell motility and compare two proliferative models: $\mathbb{P}_1$, introduced in \eqref{Macro_prol_1}, and $\mathbb{P}_2$, introduced in \eqref{Macro_prol_2}. It is worth recalling that the shared elements in these two modeling choices are: $(i)$ a proliferative rate inversely proportional to cell density; $(ii)$ an increase in proliferative activity correlating with higher levels of available oxygen; $(iii)$ the assumption that cells capabilities of moving and proliferating are inversely correlated. What distinguishes these approaches is the epigenetic or phenotypic characterization of the duality between cells' motility and proliferative dynamics. In the case of $\mathbb{P}_1$, the factor $(1-s^*)$ ensures a direct correlation between higher Snail expression and lower proliferative activity. Instead, the environmental factor has an indirect impact on the trade-off characterization, as oxygen concentration influences proliferation only indirectly by determining Snail expression. In contrast, for $\mathbb{P}_2$, the factor $1-\frac{a_2}{a_1} F\big(y^{*}, O_2(\x)\big)$ ensures that the proliferative activity decreases as $F(y^{*}, O_2(\x))$ increases, which is directly proportional to Snail expression and inversely proportional to oxygen concentration.

\begin{figure}[!h]
    \centering
    \includegraphics[width=0.7\linewidth]{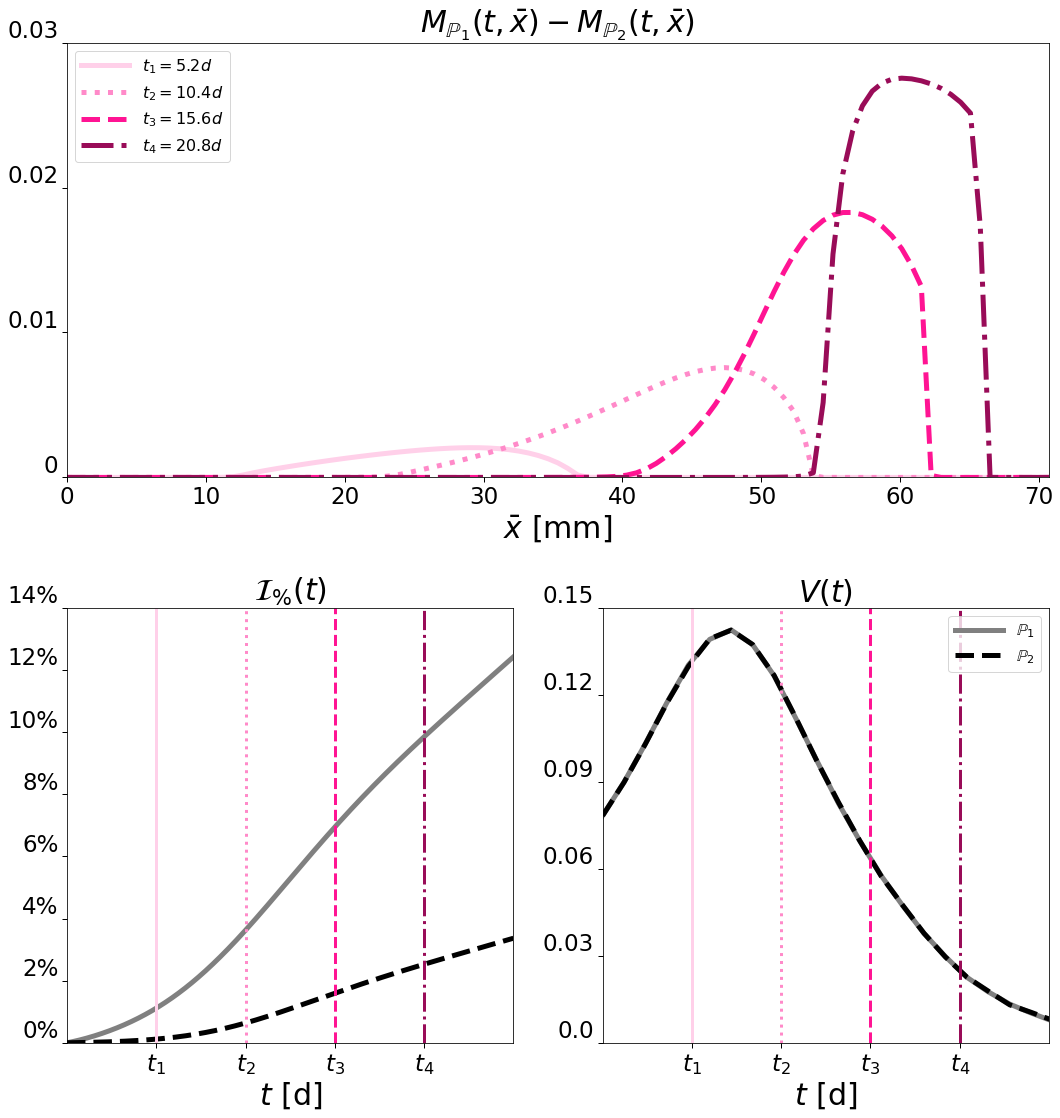}
    \caption{{\bf Experiment 2: impact of Snail expression on cell proliferation.} Top row: 1D profiles representing the evolution along the bisecting line of $\Omega$ of the difference between the solution of the tumor equation in \eqref{sim_M_ups} with the proliferation term given in \eqref{Macro_prol_1} ($M_{\mathbb{P}_1}(t,\bar{x})$) and the solution of tumor equation in \eqref{sim_M_ups} with the proliferation term given in \eqref{Macro_prol_2} ($M_{\mathbb{P}_2}(t,\bar{x})$) at four different time steps, i.e., $t=5.2$d, $t=10.4$d, $t=15.6$d, and $t=20.8$d. $\bar{x}$ indicates the spatial position along this bisecting line. Bottom row: quantification of the percentage tumor mass increment (left plot) and the velocity of the center of mass (right plot) over time with the two choices of the proliferative operator. The continuous gray line refers to the choice \eqref{Macro_prol_1}, while the dashed black line to \eqref{Macro_prol_2}. Vertical lines in the bottom-row plots mark the selected times depicted in the first row. References to the four selected time steps are repeated in each graph using consistent color and line styles. Parameter values are set as reported in Table \ref{parameter_const}.}
    \label{fig:fig2}
\end{figure}

To compare these two proliferative choices, we perform two simulations under identical environmental and initial conditions for tumor and oxygen, as well as using the same values of the model parameters. The only difference between the simulations lies in the formulation of the proliferative term. To quantify the results, we consider the 1D section of the domain $\Omega$ along the bisecting line (similar to the previous experiment) and we show the difference between the tumor densities resulting from \eqref{sim_M_ups} with $\mathbb{P}_1$ ($M_{\mathbb{P}_1}(t,\x)$) or $\mathbb{P}_2$ ($M_{\mathbb{P}_2}(t,\x)$) at four equally spaced time points: $t_1=5.2$d, $t_2=10.4$d, $t_3=15.6$d, and $t_4=20.8$d. Moreover, defining 
\begin{equation}\label{eq_tot_cell}
    Q(t) = \int\limits_{\Omega}M(t,\x) d\x
\end{equation}
as the \textit{total amount of tumor cells} in the domain $\Omega$ at time $t$, we introduce the \textit{percentage mass increment} of tumor mass from the initial configuration 
\begin{equation}\label{eq_percentage}
\mathcal{I_{\%}}(t)= 100\cdot \left(\dfrac{Q(t)-Q(0)}{Q(0)}\right)
\end{equation}
with $Q(0):=Q(t=0)$. Then, defining the \textit{center of mass} of the tumor 
$$\bm{\Upsilon}(t) = \int\limits_{\Omega} \x\bar{M}(t,\x) d\x$$
with $\bar{M}(t,\x) = M(t,\x) / Q(t)$, we compute the \textit{velocity of the center of mass}
\begin{equation}\label{eq_velocityCM}
\mathbf{V}(t) = \partial_t \bm{\Upsilon}(t)\,.
\end{equation}
The results of this experiment are shown in Figure \ref{fig:fig2}. Specifically, the top row illustrates the evolution of the difference between $M_{\mathbb{P}_1}(t,\x)$ and $M_{\mathbb{P}_2}(t,\x)$ along the 1D section. In the bottom row, we depict the temporal evolution of the mass increment $\mathcal{I}_{\%}$, defined in \eqref{eq_percentage}, (left plot), and the velocity of the center of mass, defined in \eqref{eq_velocityCM}, along the bisecting line (right plot). 

Analysis of the plot in the top row reveals that the difference is consistently positive, supporting the intuitive notion that a stronger trade-off is determined in the case of $\mathbb{P}_2$, where both the epigenetic trait and environmental factor directly contribute to module proliferation. This observation is further corroborated by the bottom-row plots of Figure \ref{fig:fig2}. In fact, we notice that the final percentage increment is six times higher for the epigenetic-driven duality ($\mathbb{P}_1$) with respect to the case in which there is a direct contribution of both epigenetic trait and environmental factor ($\mathbb{P}_2$). Considering, instead, the evolution for the center of mass, from the bottom-right plot of Figure \ref{fig:fig2} we observe a perfect overlap of its velocity dynamics throughout the experiment. This confirms the fact that any observed differences can be solely attributed to the proliferative dynamics resulting from the distinct trade-offs under analysis, while the spatial dynamics are not affected.

\subsection{Experiment 3: impact of Snail expression and hypoxia on cancer cell migration}\label{Sec_numTest3}
In this third experiment, we aim to assess the impact of Snail expression and exposure to hypoxia on cancer cell migration. Our goal is to validate our model by replicating experimental results that investigate the role of hypoxia in cell migration and determine whether motility can be triggered by inhibition or up-regulation of Snail expression. To achieve this, we specify the parameter values and the environmental conditions such that they replicate different experimental scenarios, and we compare the model outcomes with empirical observations.

We consider the parameter $g_s$, accounting for Snail transcription and, starting from the reference value of ${g_s =1.5\, (\text{molecules} \cdot \text{min}^{-1})}$, used for the experiments in Section \ref{Sec_numTest1} and \ref{Sec_numTest2}, we define up-regulation and down-regulation of Snail expression by setting as $g_s =2.1\, (\text{molecules} \cdot \text{min}^{-1})$ and $g_s =0.9\,(\text{molecules} \cdot \text{min}^{-1})$, respectively. This corresponds to variations of $0.6$ above and below the reference value. To ensure that the condition \eqref{oss_ystar} holds true in all the scenarios, for the Snail degradation rate we set  $\gamma_s = 0.03\, (\text{min}^{-1})$. Concerning the environmental conditions, by referring to \cite{mckeown2014defining} we consider levels of oxygenation compatible with normoxia ($7\%$) and pathological hypoxia ($1\%$) and we set the scaling factor $O_{2,0}$ such that, in our model, these conditions are represented by $O_2=0.7$ and $O_2=0.1$.

We aim at qualitatively replicating the experimental findings proposed in \cite{lundgren2009hypoxia} and \cite{yu2013notch1}. Specifically, in \cite{lundgren2009hypoxia} the authors investigate human breast cancer cells (cell lineage MCF-7). In their experiment, they analyze fold change in tumor cell migration by migration assays using Transwell migration chambers. Precisely, cells are suspended in upper Transwell chambers in serum-free media and allowed to migrate towards a serum gradient ($10\%$) in the lower chamber for $6$ hours. The experiment is repeated in normoxic conditions by transiently overexpressing and silencing Snail protein expression. Instead, in \cite{yu2013notch1}, the authors employ a similar methodology with the human hepatocellular carcinoma (cell lines HepG2). They assess cell motility with the same migration assays comparing experiments conducted in normoxic and hypoxic conditions.

\begin{figure}[!h]
    \centering
    \includegraphics[width=0.65\linewidth]{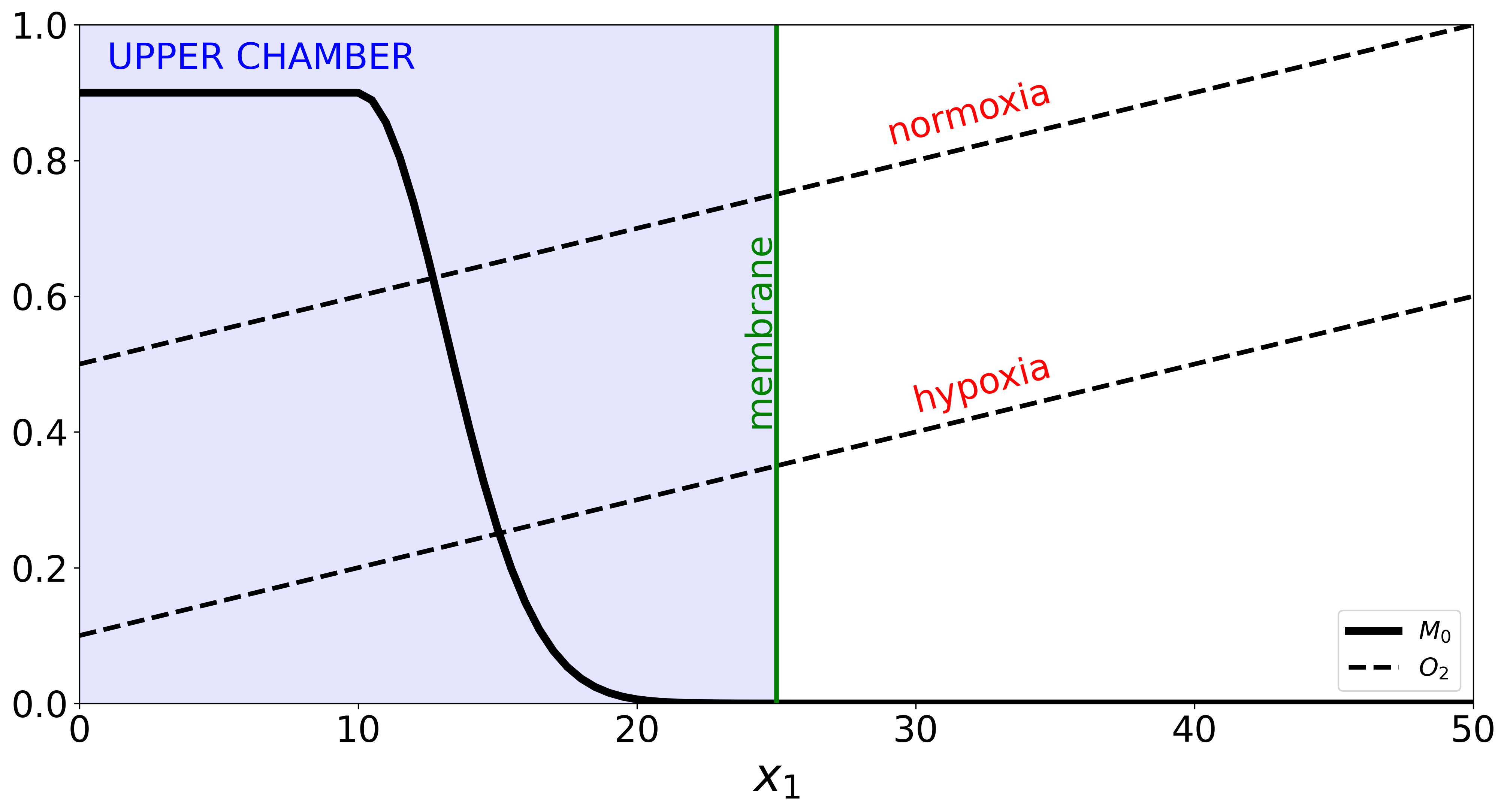}
    \caption{{\bf Experiment 3: initial conditions and setting.} 1D graphical illustration of the setting implemented for studying the impact of Snail expression and hypoxia on cancer cell migration. The domain $\Omega$ is divided into two chambers. Tumor cells are distributed in the upper chamber, i.e., in $\Omega_U=[0,25]\times[0,50]\,\text{mm}^2$, accordingly to \eqref{M_IC_chamber}, while no cells are initially located in the bottom chamber, i.e., in $\Omega_B=[25,50]\times[0,50]\,\text{mm}^2$. Two different linear oxygen distributions (for normoxia and hypoxia scenarios) are represented as dashed black lines. The central membrane dividing the two chambers is shown in green.}
    \label{fig:fig3_exp}
\end{figure}

We replicate the chamber setup by considering our square domain $\Omega$ intersected by a vertical membrane parallel to the $x_2$ axis and positioned at $x_1=25\,\text{mm}$. A 1D section of the chamber setup is illustrated in Figure \ref{fig:fig3_exp}. 
Here, the left part of the domain (for $x_1<25$ mm) represents the upper chamber, where all cells are initially distributed following
\begin{equation}\label{M_IC_chamber}
M^0(\x)= 
    \begin{cases}
      I_M \qquad\qquad\qquad\qquad x_1 < x_s\\[0.3cm]
      I_Me^{-\dfrac{(x_1-x_s)^2}{\theta^2_M}} \qquad x_1 \geq x_s
    \end{cases}
\end{equation}
where $I_M = 0.9$, $x_s = 10\, \text{mm} $ and $\theta_M = 3\, \text{mm}$. We consider the following linear expression for the oxygen distribution
\[
O(\x) = O_{min} + \frac{(O_{max} - O_{min})}{50} x_1
\]
with $O_{min}=0.7$ and $O_{max}=1.0$ in normoxic conditions, while $O_{min}=0.1$ and $O_{max}=0.4$ in hypoxic conditions. This choice establishes a fixed oxygen gradient along the chamber, which is consistent with the biological setting. We conduct five experiments, which are summarized in Table \ref{sum_experiment}.

\begin{table}[!h]
\begin{center}
     \begin{tabular}{|c|c|c|}
   \toprule  
  \rule{0pt}{1.5ex} Name & \myalign{c|}{Oxygenation} & \myalign{c|}{Snail expression}\\
  \midrule
   \rule{0pt}{3ex} A & normoxia &  control  \\[0.3ex]
    \rule{0pt}{3ex} B & normoxia &  up-regulated  \\[0.3ex]
    \rule{0pt}{3ex} C & normoxia &  down-regulated \\[0.3ex]
     \rule{0pt}{3ex} D & hypoxia &  control  \\[0.3ex]
      \rule{0pt}{3ex} E & hypoxia &  down-regulated  \\[0.3ex]
\bottomrule
    \end{tabular}
\caption{{\bf Summary of the conducted experiments.} The table shows the information regarding tissue oxygenation and Snail expression in the five scenarios analyzed in Section \ref{Sec_numTest3}.}
\label{sum_experiment}
\end{center}
\end{table}
\noindent Under the aforementioned conditions, we allow cells to move in response to the environmental stimuli for a duration of $T=6$ hours. Subsequently, we measure the quantity of tumor mass that has passed through the membrane as
\[
\tilde{Q}(t) = \int\limits_{{\Omega_B}}M(t,\x) d\x\,,
\]
where ${\Omega_B} = [25,50] \times [0,50]\, \text{mm}^2$ represents the bottom chamber.

\begin{figure}[!h]
    \centering
    \includegraphics[width=0.45\linewidth]{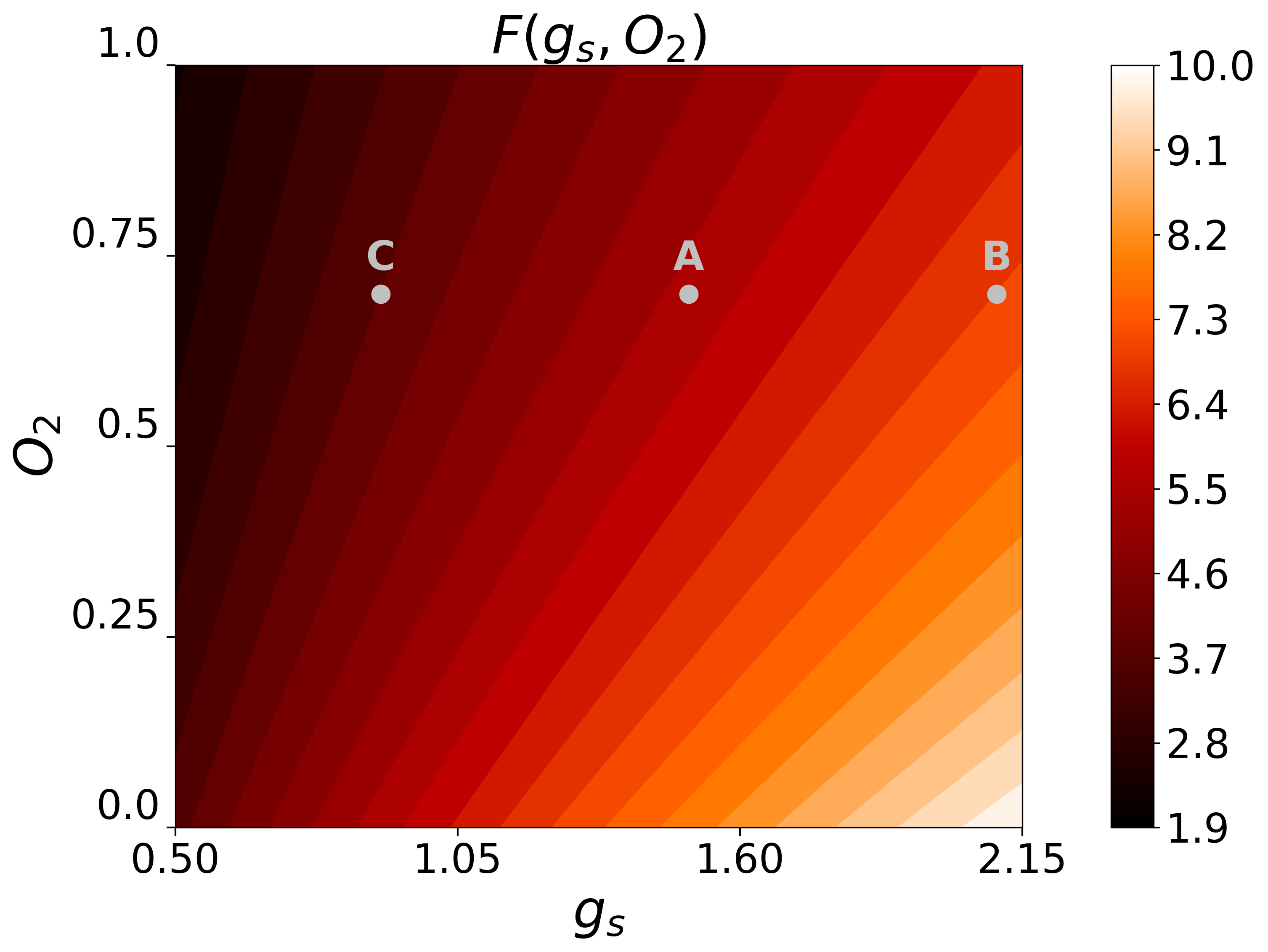}
    \hfill
    \includegraphics[width=0.45\linewidth]{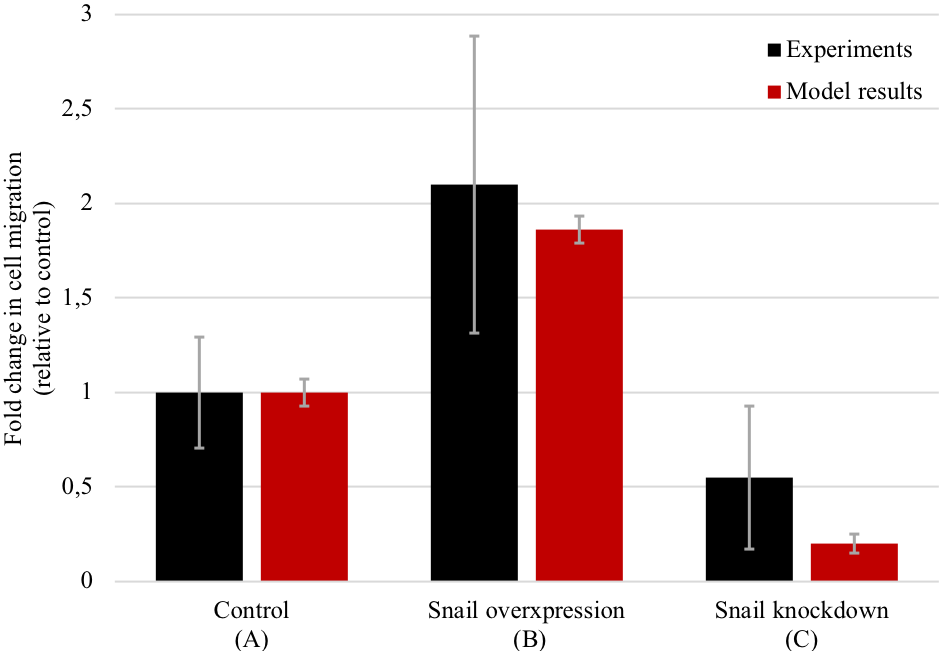}\vspace{0.5cm}
    \includegraphics[width=0.45\linewidth]{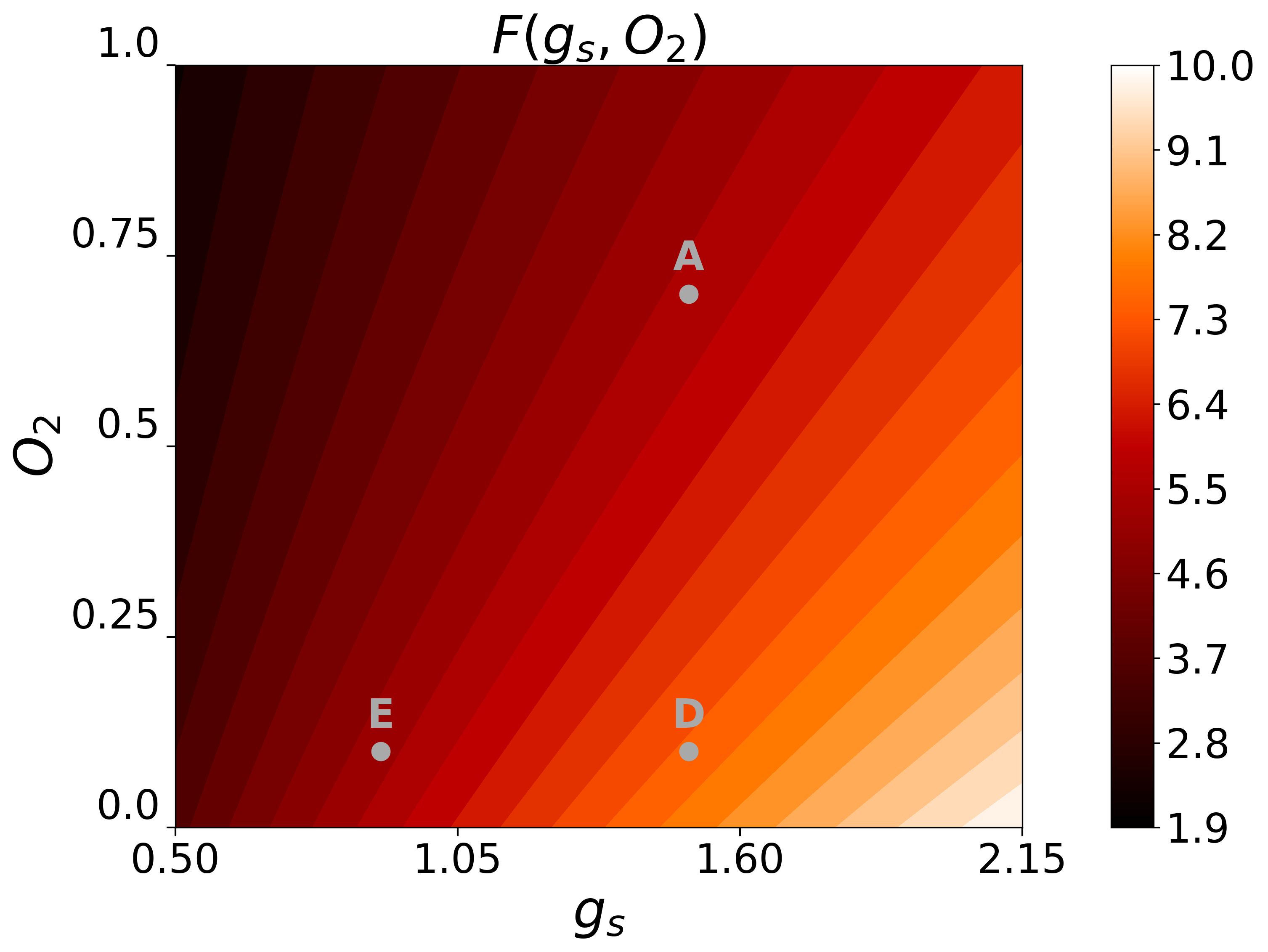}
    \hfill
    \includegraphics[width=0.45\linewidth]{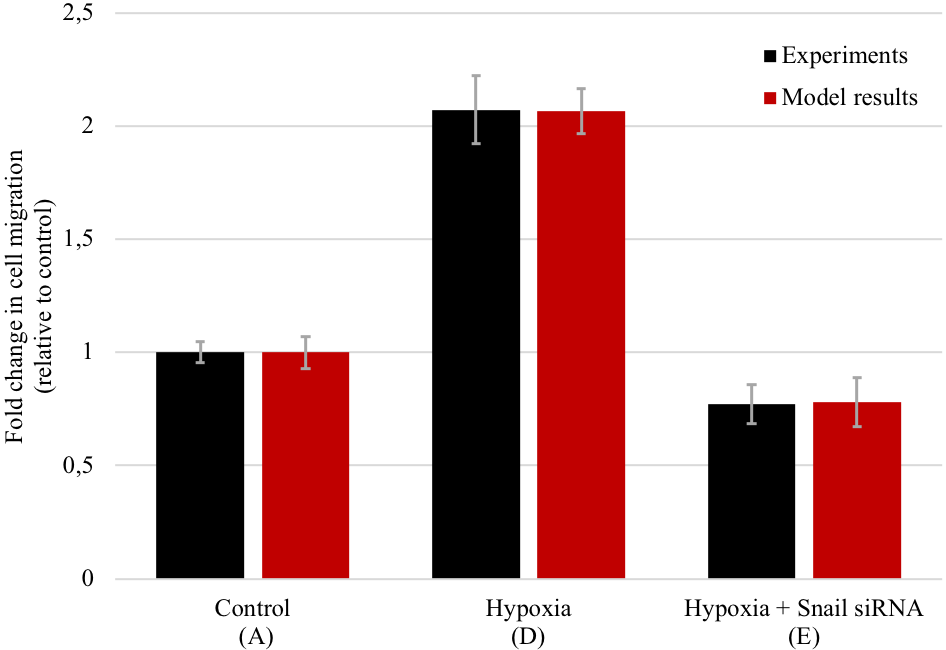}
    \caption{{\bf Experiment 3: impact of Snail expression and hypoxia on cancer cell migration.} Left: 
     representation of the level curve of the tactic sensitivity $F$ with respect to the Snail transcription rate $g_s$ and the oxygen concentration $O_2$. The parameter combinations referring to the five analyzed scenarios are indicated with gray bullets. Right: comparison between {\it in-vitro} (black) and {\it in-silico} (red) results concerning the fold change in cell migration (relative to control) in the five different scenarios. Precisely, in the top-right panel MCF-7 cells are considered in normal conditions (control, scenario A), with Snail overexpression (B), or with Snail knockdown (C). {\it In-vitro} data are obtained from Figure 3B and 3D in \cite{lundgren2009hypoxia}. In the bottom-right panel, HepG2 cells are considered in normoxic conditions (control, A), hypoxic conditions (D), or hypoxic conditions with Snail silencing, indicated as Snail siRNA (E). {\it In-vitro} data are obtained from Figure 1A and 1E in \cite{yu2013notch1}. In both cases, {\it in-silico} results are obtained by simulating system \eqref{sim_M_ups} under normoxic and hypoxic conditions and for different values of the parameter $g_s$. Means $\pm$ std in the simulations are obtained by varying the value of the parameter $g_s$ within a range of $\pm 0.05$.}
    \label{fig:fig3}
\end{figure}

We designate the results obtained in scenario (A) as the control case and we use them to normalize the outcomes of the other experiments. Figure \ref{fig:fig3} collects all the results of the five tests. In the top row, we show the results related to the scenarios (A), (B), and (C) and we compare them with the data taken from \cite{lundgren2009hypoxia}, while in the bottom row, we refer to scenarios (A), (D), and (E) and we compare the results with the data taken from \cite{yu2013notch1}. Each row comprises two columns. The left column provides a map of the values taken by $F$, given in \eqref{F}, as a function of levels of oxygenation ($O_2$) and Snail transcription ($g_s$), while the right column provides histograms comparing the results of \textit{in-vitro} (black) and \textit{in-silico} (red) experiments. \\

\noindent As clearly shown in Figure \ref{fig:fig3}, for both the breast cancer and the hepatocarcinoma cases, the model is able to effectively replicate the experimental data. Specifically, in the case of breast cancer, the \textit{in-silico} results closely resemble those obtained in the \textit{in-vitro} experiments for the scenario (B). In scenario (C), any discrepancy appears to be merely apparent, as the \textit{in-silico} results fall within the error band of the \textit{in-vitro} experiments, which notably exhibits a wider range of data. For hepatocarcinoma, there is a remarkably high correspondence between the \textit{in-vitro} and \textit{in-silico} data, and experiments (D) and (E) show a notable match. Furthermore, it is worth noticing how the level curves of $F(g_s,O_2)$ (left column of Figure \ref{fig:fig3}) provide insight into the experimental observations. Specifically, previous experimental works have noticed that the knockdown of Snail nullifies the motility advantage gained under hypoxic conditions (D) compared to normoxia (A), bringing the motility to a level comparable with the control case (as observable by comparing scenarios (E) and (A)). In our case, this can be observed a priori by looking at the location of the corresponding bullets on the level plot of $F(g_s,O_2)$: (A) and (E) are located, in fact, almost on the same level curve. This implies that the term governing cell movement assumes comparable values in both experiments (slightly lower in (E))  when considering equal cell density and a consistent oxygen gradient, which is maintained at the same value across the domain and for all the experiments. Therefore, comparable levels of fold changes in cell migration are expected as well as the slight discrepancy in the number of cells passing through the membrane between the two cases.

\subsection{Experiment 4: hypoxia-driven ring structure in tumor and Snail distributions}\label{Sec_numTest4}
As the last experiment, we refer to additional results shown in \cite{lundgren2009hypoxia}, where the authors use a model system of hypoxia {\it in-vivo} to investigate the expression of Snail in non-invasive ductal carcinoma in situ (DCIS), namely an early breast cancer. Considering a central necrotic area, their analysis of several DCIS samples revealed a typical pattern of HIF-1$\alpha$ expression, with increasing staining intensity approaching the areas of necrosis, and similar spatial distribution for the nuclear expression of Snail, gradually increasing when approaching the necrosis (see Figure 6 in \cite{lundgren2009hypoxia}). In particular, the authors show that hypoxia induces Snail expression independently of other EMT markers. 

\begin{figure}[!h]
    \centering
    \includegraphics[width=.8\linewidth]{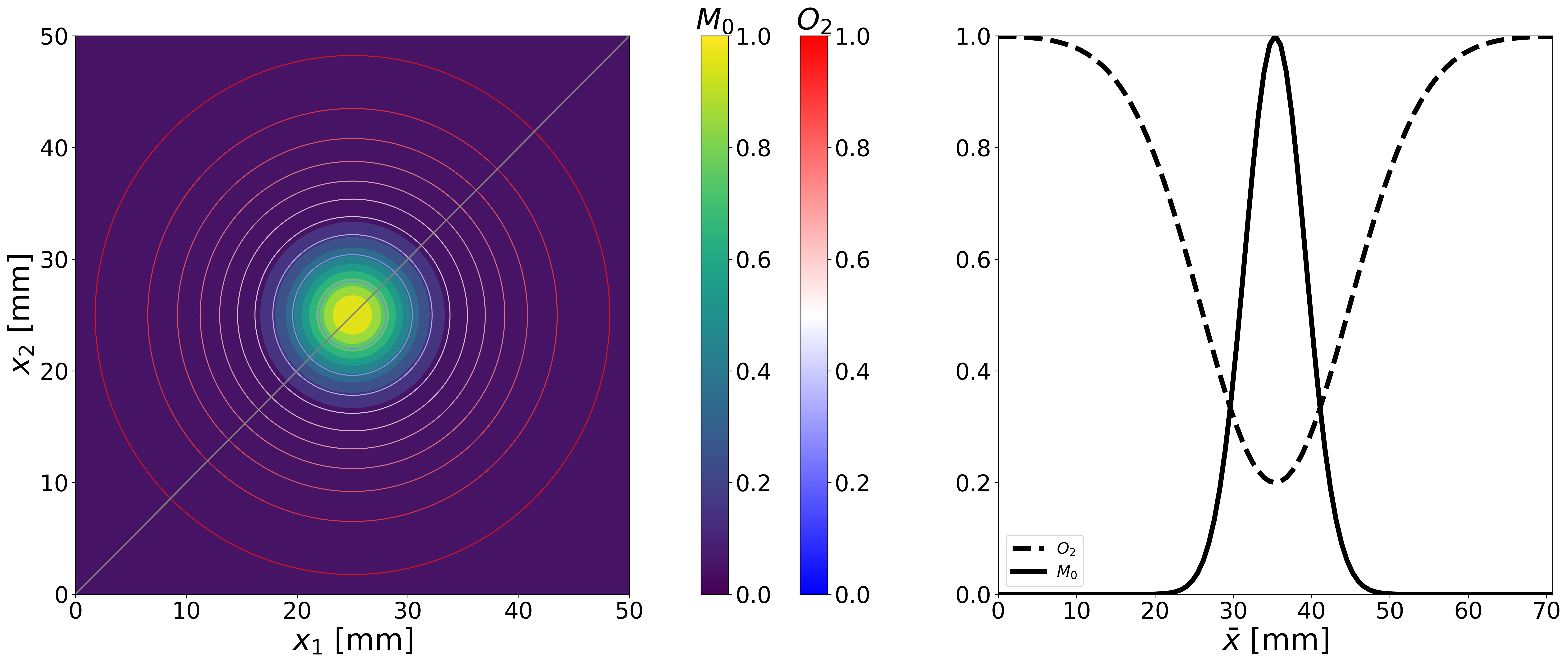}
    \caption{{\bf Experiment 4: initial conditions.} Left: initial Gaussian distribution of the tumor cells $M^0$, located at position $\mathbf{x}_M = [25,25]\,\text{mm}$ in the domain $\Omega=[0,50]\times[0,50]\,\text{mm}^2$, together with the level plot for the fixed Gaussian distribution of oxygen $O_2$, located at the same position. Right: 1D profiles of tumor (continuous line) and oxygen (dashed line) distribution along the bisecting line (light gray line in the 2D plot) of the domain $\Omega$. $\bar{x}$ indicates the spatial position along this bisecting line.}
    \label{fig:fig4_ic}
\end{figure}

To qualitatively reproduce these observations, we simulate an initial tumor mass located in the center of the domain. We assume an oxygen distribution that decreases towards the center of the domain, leading to highly hypoxic (or necrotic) areas in regions with higher cell density. The initial condition for cancer cells is given by \eqref{eq_inM} setting $I_M=1$, $\mathbf{x}_M = [25,25]\, \text{mm}$, and $\theta_M=5\, \text{mm}$, while for the fixed oxygen distribution we consider $1-O(\mathbf{x})$, with $O(\x)$ defined by \eqref{eq_inO} and $I_{O_2}=0.8$, $\mathbf{x}_{O_2} = [25,25]\, \text{mm}$, and $\theta_{O_2}=13\, \text{mm}$. The initial conditions for tumor cells and oxygen are illustrated in Figure \ref{fig:fig4_ic}.

\begin{figure}[!h]
    \centering
    \includegraphics[width=1\linewidth]{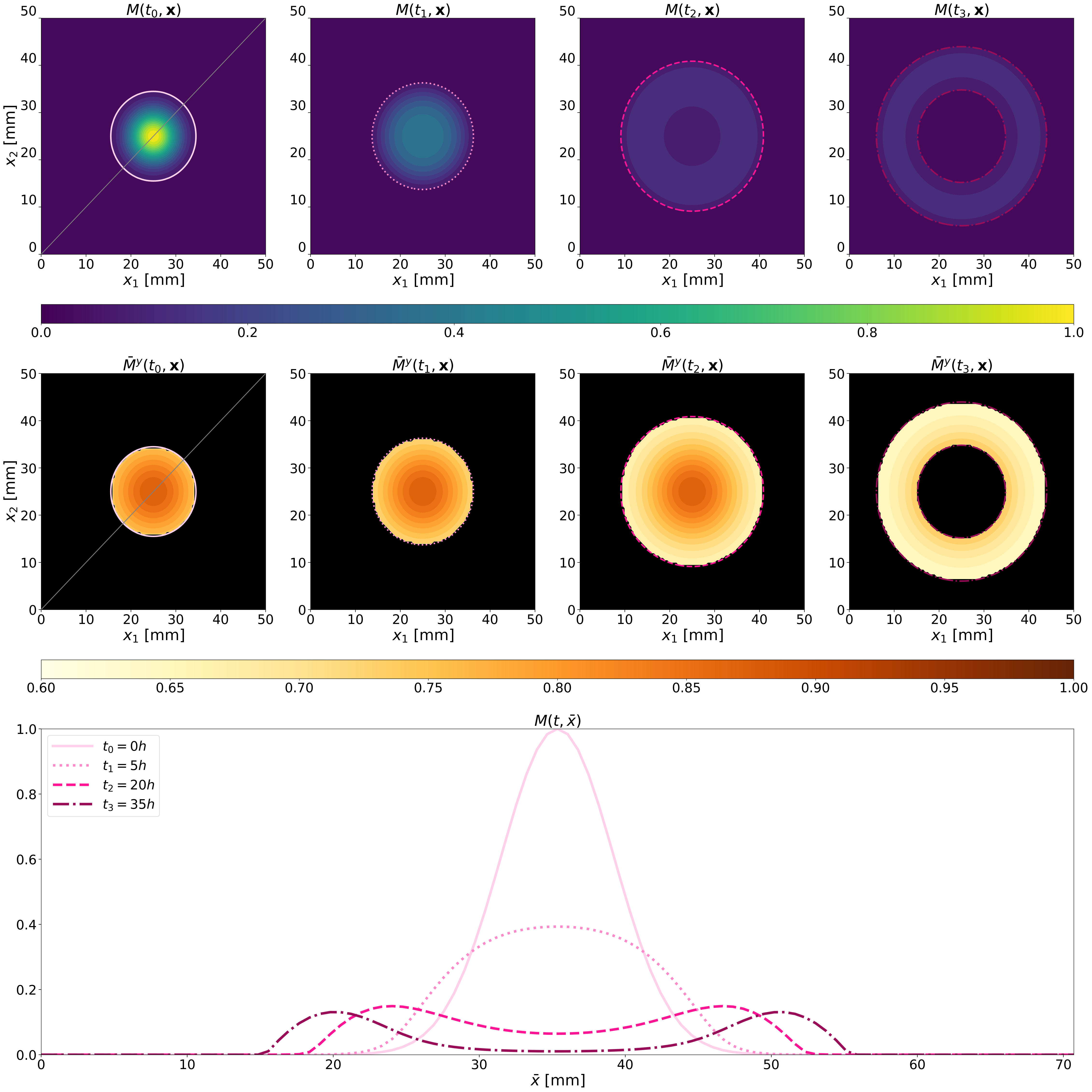}
    \caption{{\bf Experiment 4: hypoxia-driven ring structure in tumor and Snail distribution.} Evolution of model \eqref{sim_M_ups} with the initial conditions shown in Figure \ref{fig:fig4_ic}. Top row: evolution of the tumor mass $M(t,\x)$ in the domain $\Omega=[0,50]\times[0,50]\,\text{mm}^2$ at four different time step, i.e., initial time $t_0=0$h (continuous light pink line), and progression at $t_1=5$h (dot pink line), $t_2=20$h (dashed dark pink line), and $t_3=35$h (dot-dashed purple line). Middle row: evolution of $\bar{M}^y(t,\x)$ defined in \eqref{MY_test4} at the same time steps. Bottom row: 1D profiles of the tumor mass evolution along the bisecting line (light gray line in the first 2D plot) of the domain $\Omega$ at the same time steps. $\bar{x}$ indicates the spatial position along this bisecting line. References to the four selected time steps are repeated in each graph using consistent color and line styles. Parameter values are set as reported in Table \ref{parameter_const}.}
    \label{fig:fig4}
\end{figure}

\noindent Figure \ref{fig:fig4} collects the results of this experiment at four time steps: $t_0=0$h, $t_1=5$h, $t_2=20$h, and $t_3=35$h. The first row depicts a 2D representation of the tumor mass, including density map and contour lines highlighting the tumor's edge (defined by a density threshold corresponding to $10\%$ of the carrying capacity). Then, defined the average expression of the Snail protein in the cell population as
\begin{equation}\label{MY_test4}
    \bar{M}^y(t,\x) = \dfrac{M^y(t,\x)}{M(t,\x)}= \dfrac{a_2}{a_1} F(y^{*},O_2)\quad \text{for } \x\in\text{supp}(M)\,,
\end{equation}
we illustrate its evolution in the second row of Figure \ref{fig:fig4}. Finally, in the third row, the 1D section of the tumor mass density along the bisecting line at the four specified time steps is shown.
\noindent We observe that, initially, both anti-crowding and chemotactic stimuli point in the same direction, which tends to quickly move the cells away from the central hypoxic region, where cell density is high and oxygen concentration is low. Consequently, in this initial phase, cell migration is rapid, leading to a fast transformation of the peaked initial Gaussian for cell distribution into a smoother bubble profile (as shown at time $t_1$). As time progresses, the prevalence of chemotactic motion results in a depletion of cells from the central mass, gradually giving rise to a ring-like structure (times $t_2$ and $t_3$). During this phase, movement starts to slow down due to two main factors. Firstly, comparing the position of the ring with respect to the oxygen profile (shown in Figure \ref{fig:fig4_ic}), we observe a decrease in motility caused by both high levels of oxygenation (reducing the tactic sensitivity $F(y^{*},O_2)$) and low oxygen gradients (reducing the chemotactic stimulus). Secondly, the slowdown is due to the anti-crowding mechanism, which, once the void forms at the center of the mass, would induce cells from the inner part of the ring to move towards the center, conflicting with the chemoattractant-driven movement. These observations are also consistent with the plots in the second row of Figure \ref{fig:fig4}, where $\bar{M}^y(t,\x)$ is shown. They illustrate how, initially, the average expression of Snail is high, inducing rapid cell migration, and it increases approaching the central hypoxic core. Then, while the tumor mass moves outward, this expression decreases as cells reach more oxygenated areas, still maintaining higher values around the inner border of the ring. It is interesting how the model is able to qualitatively capture the two main dynamics shown in the data from \cite{lundgren2009hypoxia}. In fact, the model reproduces both the experimentally observed ring shape of the tumor mass and the spatial distribution of the average Snail expression, mirroring the findings of the experimental study.

\section{Conclusion}\label{sec_conclusion}

The migration of tumor cells in response to oxygen concentration gradients remains a critical area of research in cancer biology. While the role of hypoxia in promoting tumor aggressiveness and metastasis is well recognized, the exact mechanisms driving cell migration in response to oxygen levels are still an area of investigation, and understanding these mechanisms may be crucial for developing effective therapeutic strategies.

In this study, we developed a novel mathematical model to investigate the interplay between hypoxia, molecular signaling pathways, and tumor cell migration. Specifically, we proposed a multi-scale model that naturally integrates single-cell behavior driven by Snail expression with macroscopic scale dynamics describing tumor migration in the tissue. Our approach employs tools and methods from the kinetic theory of active particles to construct a kinetic transport equation that describes the evolution of the tumor cell distribution based on detailed microscopic dynamics. By employing proper scaling arguments, we formally derived the equations for the statistical moments of the cell distribution. These capture cell density dynamics, influenced by limited non-linear diffusion and oxygen-mediated drift, and the evolution of the average Snail expression within the tumor population, which directly relates to tumor migratory capability. Overall, our model offers a detailed description of macroscopic tumor cell dynamics, considering the effect of microscopic Snail signaling pathways in the mechanisms of tumor response to hypoxia. This modeling approach represents a promising way to integrate molecular signaling pathways with cell migration dynamics.

We validated the model in different scenarios with biological relevance, focusing on the role of chemotactic-driven motion and anti-crowding effects, and analyzing the effect of Snail expression on cell migration and proliferation. We showed the reliability of our approach by testing its ability to quantitatively replicate experimental results from two different studies published in the literature. Precisely, we investigated the effect of hypoxia and Snail knockdown on the motility of cancer cells, comparing our results with those presented by \cite{yu2013notch1} on human hepatocarcinomas. Moreover, we considered the findings in \cite{lundgren2009hypoxia} and we replicate \textit{in-silico} the results regarding the effect of Snail over-expression and Snail knockdown on the migration capability of human breast cancer cells in normoxic conditions. We also analyzed the spatial distribution of Snail expression within the tumor mass in response to hypoxia, showing how the model is able to reproduce the spatial patterns experimentally observed in \cite{lundgren2009hypoxia}. These results support the idea that our mathematical framework can offer new perspectives for interpreting experimental data and understanding the underlying biological mechanisms driving tumor migration.

Moving forward, it will be important to explore the implications of our findings in the context of clinical outcomes and therapeutic interventions. Particularly, our results highlight the importance of considering the dynamic regulation of Snail expression in response to hypoxia. This finding underscores the potential significance of developing strategies to target Snail as a therapeutic approach to control tumor cell migration and metastasis. Furthermore, incorporating heterogeneous and dynamic environmental factors, such as a non-stationary oxygen distribution, could improve the predictive power of our model and enhance the quantitative fitting of the experimental data, ultimately leading to a better understanding of tumor invasion.

In summary, the proposed mathematical modeling approach is a novel and valuable tool to integrate detailed descriptions of microscopic cell dynamics with cell evolution at a macroscopic (tissue) level. In particular, the multi-scale modeling approach allows to properly derive the macroscopic terms driving cell evolution from a detailed description of the single-cell dynamics, instead of phenomenological stating them directly at the macroscopic level. Our findings offer interesting interpretations of the complex dynamics underlying tumor progression and motility, providing new perspectives for interpreting experimental data and understanding the biological mechanisms driving tumor development. This also paves the way for personalized medicine approaches tailored to individual tumor characteristics.

\subsection*{Acknowledgement}
This paper has been partially supported by the Italian Ministry of Education, Universities and Research, through the MIUR grant Dipartimento di Eccellenza 2018-2022, project E11G18000350001 (MC, GC, MD), by the National Group of Mathematical Physics (GNFM-INdAM) through the INdAM–GNFM Project (CUP E53C22001930001) "From kinetic to macroscopic models for tumor-immune system competition" (MC). This work has been partially supported by the State Research Agency of the Spanish Ministry of Science and FEDER-EU, project PID2022-137228OB-I00 (MICIU/AEI /10.13039/501100011033); by Modeling Nature Research Unit, Grant QUAL21-011 funded by Consejería de Universidad, Investigaci\'on e Innovaci\'on (Junta de Andalucía) (MC). MC also acknowledges support from City of Hope’s Global Scholar Program.

\subsection*{Author contribution statement}
Conceptualization: G.C., M.C., M.D. Methodology: G.C., M.C., M.D. Formal analysis: G.C., M.C., M.D. Resources: G.C., M.C., M.D. Software: G.C. Visualization: G.C., M.C. Writing-original draft preparation: G.C., M.C., M.D.  Writing—review and editing: G.C., M.C., M.D. Supervision: M.D. Project administration: M.D. Funding acquisition: M.C., M.D. All authors contributed to the article and approved the submitted version.

\subsection*{Competing interests}
The authors declare that the research was conducted in the absence of any commercial or financial relationships that could be construed as a potential conflict of interest.

\newcommand{\noopsort}[1]{}
\nocite{*}
\bibliographystyle{abbrv}
\bibliography{Biblio}

\end{document}